\documentclass[10pt]{iopart}
\usepackage{graphicx}
\usepackage{ulem}
\begin{document}

\title[Experimental and numerical study of error fields in the CNT 
       stellarator]{Experimental and numerical study of error fields in the 
       CNT stellarator}

\author{K C Hammond$^1$, A Anichowski$^1$, P W Brenner$^1$, 
        T S Pedersen$^{1,2}$, S Raftopoulos$^{3}$, P Traverso$^{1,4}$, 
        F A Volpe$^1$}

\address{$^1$ Dept.~of Applied Physics and Applied Mathematics, 
         Columbia University, New York, NY 10027, USA}
\address{$^2$ [Present institution: Max Planck Institute for Plasma Physics, 
         17491 Greifswald, Germany]}
\address{$^3$ Princeton Plasma Physics Laboratory, Princeton, NJ 08543, USA}
\address{$^4$ [Present institution: Dept.~of Physics, Auburn University, 
         Auburn, AL 36849, USA]}

\begin{abstract}
Sources of error fields were indirectly inferred in a stellarator by reconciling
computed and numerical flux surfaces. Sources considered so far include the 
displacements and tilts of the four circular coils featured in the simple CNT 
stellarator. The flux surfaces were measured by means of an electron beam and 
fluorescent rod, and were computed by means of a Biot-Savart field-line tracing 
code. If the ideal coil locations and orientations are used in the computation, 
agreement with measurements is poor. Discrepancies are ascribed to errors in 
the positioning and orientation of the in-vessel interlocked coils. To that 
end, an iterative numerical method was developed. A Newton-Raphson algorithm 
searches for the coils' displacements and tilts that minimize the discrepancy 
between the measured and computed flux surfaces. This method was verified by 
misplacing and tilting the coils in a numerical model of CNT, calculating the 
flux surfaces that they generated, and testing the algorithm's ability to 
deduce the coils' displacements and tilts. Subsequently, the numerical method 
was applied to the experimental data, arriving at a set of coil displacements 
whose resulting field errors exhibited significantly improved agreement with 
the experimental results. 
\end{abstract}

\maketitle
\ioptwocol

\section{Introduction}
\label{sect:introduction}

Error fields (EFs) have been and are the subject of intense study in tokamaks 
\cite{reimerdes2011,hender2007} where relative errors as small as $10^{-4}$ or 
even $10^{-5}$ are known to affect stability, cause disruptions, and degrade 
confinement and plasma rotation. Error fields are obviously also very important 
in modern transport-optimized stellarators \cite{mynick2006,xanthopoulos2014}, 
whose performances rely on carefully optimized 3D magnetic fields. 

The specially shaped coils that generate such fields are numerically optimized 
and, typically, are built and positioned with very high precision. In other 
words, errors are minimized at the construction stage. Vacuum fields are then 
experimentally characterized by a standard technique involving an electron beam 
and a fluorescent rod 
\cite{sinclair1970,colchin1989,jaenicke1993,pedersen_pop2006}. 
Incidentally, it should be noted that, since this technique requires flux 
surfaces to exist in a vacuum, it is not applicable to devices such as tokamaks 
that require plasma current to generate flux surfaces. The measured surfaces 
are usually confirmed to be in good agreement with the desired, optimal 
configuration. However, if not in agreement, a comparison of computed and 
measured flux surfaces can shed light on possible sources of errors. At that 
point one can either (1) correct the error ``at the source'' (reposition one or 
more coils) \cite{colchin_iaea1989} or (2) apply EF corrections by means of a 
separate set of dedicated coils. This latter approach is quite common in 
tokamaks \cite{reimerdes2011,hender2007,otte2003}. As for stellarators, EF 
corrections were applied in LHD by 
means of Resonant Magnetic Perturbation (RMP) coils, also deployed in other MHD 
studies \cite{sakakibara2013}. W7-AS used ``special'' and ``control'' coils to 
vary the toroidal mirror term and boundary island geometry and study their 
effect on plasma properties \cite{hirsch2008}, and similar uses are envisioned 
for the ``trim coils'' in W7-X \cite{rummel2012}. Earlier possibilities for EF 
correction in W7-X were discussed in Ref.~\cite{kisslinger2005}.

Stellarators are considered complicated to build. It has been suggested that 
their attractiveness as reactors could increase if their construction is 
simplified, or construction tolerance relaxed, without significantly degrading 
the plasma properties \cite{landreman2016}. At this point it is highly 
hypothetical, but a possible route to simplification could consist in: 
i) slightly relaxing the tolerance and ii) correcting the errors \textit{a 
posteriori}, either by approach (1) or (2) listed above.
In either approach, it is useful to experimentally quantify the displacements, 
tilts and, possibly, deformations of the actual coils, compared with design 
values. This is useful anyway, even if stricter tolerance is adopted, to 
confirm that the construction imperfections are indeed smaller. 

As a first step in exploring such route, we have complemented the 
well-established experimental technique mentioned above with metrology 
measurements and with a numerical method that ``inverts'' the flux surface 
errors into error sources such as imperfections in coils' positions and tilts.

\begin{figure}
    \begin{center}
    \includegraphics[width=0.425\textwidth]{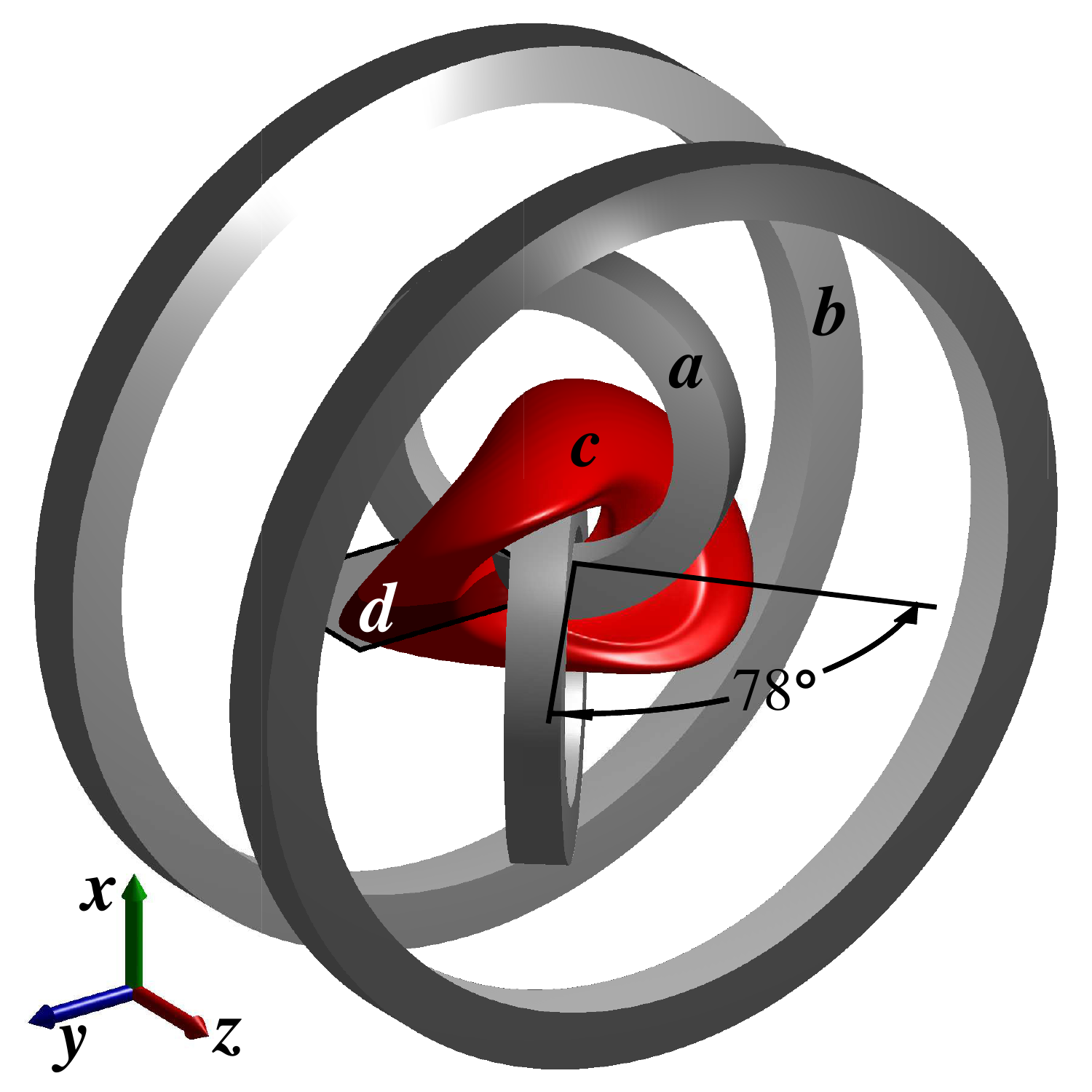}
    \caption{Schematic of the CNT coil configuration. (a) interlocked (IL)
             coils, (b) poloidal field (PF) coils, (c) the 
             last closed flux surface, (d) plane at $\phi = 90^\circ$ at 
             which Poincar\'{e} cross-sections are measured as described in 
             Section \ref{sect:flux_surf_meas}.}
    \label{fig:coil_schematic}
    \end{center}
\end{figure}

The CNT stellarator, constructed in 2004 \cite{pedersen2004}, is notable for
its simple coil configuration and for having the lowest aspect ratio 
of any stellarator ever built \cite{pedersen_pop2006}. The magnetic field is 
generated by a system of four planar, circular coils: two interlocked (IL) 
coils and two poloidal field (PF) coils, the latter of which form a 
Helmholtz pair (Fig.~\ref{fig:coil_schematic}). This configuration, as 
designed, generates a set of 
toroidal nested closed flux surfaces with two field-periods. The 
radial profile of rotational transform \sout{$\iota$}, as well as the 
general shape of the flux surfaces, can be controlled by varying two main 
parameters: the tilt angle between the IL coils and the ratio of the current in 
the IL coils to the current in the PF coils, $I_{IL}/I_{PF}$. CNT was 
constructed to permit 
three different IL coil tilt angles $\theta_{tilt}$: $64^\circ$, $78^\circ$, 
and $88^\circ$. These 
angles were chosen for their distinct \sout{$\iota$} profiles as well as for 
their resiliency to EFs \cite{pedersen2004}.

Error field resiliency was an especially important consideration for CNT's 
coil configurations due to the permissive tolerances used in the 
construction of the coils and the vacuum vessel. Whereas present-day 
stellarators are typically built to tolerances on the order of $10^{-3}$ to 
$10^{-4}$ \cite{imagawa1998,wanner2000}, CNT's tolerances were of order 
$10^{-2}$ to $10^{-3}$ to minimize the cost and complexity of construction. 

From 2005-2010, the interlocked coils were kept in the $\theta_{tilt} 
= 64^\circ$ configuration. The flux surfaces for this configuration were 
measured experimentally and found to agree very well with numerical predictions
\cite{pedersen_pop2006}. More recently, CNT's coils were switched to the 
$\theta_{tilt} = 78^\circ$ configuration. This configuration is predicted to 
have less magnetic shear than the previous one. As a result, EFs that 
resonate with rational surfaces within the \sout{$\iota$} profile are expected 
to have more significant effects on the magnetic geometry, leading to larger 
islands and equilibrium deformation.

A detailed understanding of the magnetic geometry in CNT is a central objective
of the CNT research program. It will allow for more precise alignment of 
diagnostics and improve the accuracy of equilibrium modeling and reconstruction.
It will also assist in the design of equipment whose shaping and placement 
are heavily dependent on the field, such as island divertors.

In this paper we present the first detailed measurements of the flux surface 
geometry of the $\theta_{tilt} = 78^\circ$ configuration, which exhibits
noticeable disagreements with predictions (Sec.~\ref{sect:flux_surf_meas}). 
We then describe efforts to diagnose the sources of the 
EFs under the assumption that the main sources of error are 
displacements of the coils from their design positions. These efforts include
(1) photogrammetric measurements of the positions of
the PF coils (Sec.~\ref{sect:photogrammetry}), 
(2) studies of the effects of different classes of displacements on the 
rotational transform \sout{$\iota$} (Sec.~\ref{sect:numerical_studies}), 
and (3) design of an optimization algorithm to find the most 
likely displacements of the IL coils that lead to the observed flux surface 
geometry (Sec.~\ref{sect:optimization}). We then discuss the potential 
broader applicability of the optimization algorithm to more complex coil 
configurations (Sec.~\ref{sect:discussion}) and briefly describe future work 
(Sec.~\ref{sect:summary}).

\section{Flux Surface Measurments}
\label{sect:flux_surf_meas}

\subsection{Experimental setup}
\label{subsect:exp_setup}

The flux surface geometry in CNT is measured with a standard technique involving
an electron beam and a fluorescent rod \cite{pedersen_pop2006,jaenicke1993}.
Electrons are emitted from an electron gun 
at energies of roughly 80 eV. The electrons travel along a 
field line until they strike an obstacle, either a part of the vessel (if 
an open field line) or an 
aluminum rod that extends into the confining region. The aluminum rod is 
coated with ZnO:Zn fluorescent powder that emits blue-green light (505 nm) when 
struck by electrons. In CNT, two fluorescent rods are positioned 
at the mid-plane of the vacuum vessel, which coincides with the toroidal 
cross-section at $\phi = 90^\circ$ (Fig.~\ref{fig:coil_schematic}). The rods 
can be rotated across this surface
with an external actuation mechanism described in Ref.~\cite{pedersen_pop2006}.
Background pressures are maintained at the base
pressure of the CNT vacuum vessel ($<10^{-8}$ Torr) to minimize collisions with
neutral atoms and molecules.

If a long-exposure photograph is taken of the fluorescent rods as they 
are scanned across the $\phi = 90^\circ$ plane during the emission of an 
electron beam, the resulting image will show a cross-section of the flux surface
on which the beam was emitted. Certain parts of the cross-section may not 
appear in the image due to shadowing by the electron gun or the limited extent
of the fluorescent rods. For the measurements described in this paper, a 
digital camera was positioned outside the vacuum vessel to face the plane of 
the cross-section through a fused-silica viewport.  Images were acquired using 
ten seconds of exposure time. In addition to the fluorescent glow from the 
rods, the images contained some regions of stray light.
This was due to blackbody emission from the electron gun filament reflecting 
off components in the vacuum vessel.
Regions of stray light were eliminated manually from each image before 
processing.

The relative positions of each pixel were determined based on the length of the
rod in the image. The absolute positions of each pixel were then determined by
comparing the position of the rod to the position of a fixed landmark in the 
chamber. The lens axis of the camera was 
confirmed to be within one degree of perpendicular to the $\phi = 90^\circ$
plane through the use of a leveling tool. This uncertainty could contribute
to up to 3 mm of displacement of the resulting image, as the camera was placed 
roughly one meter away from the cross-section.

\subsection{Results}
\label{subsect:exp_results}

\begin{figure}
    \begin{center}
    \includegraphics[width=0.5\textwidth]{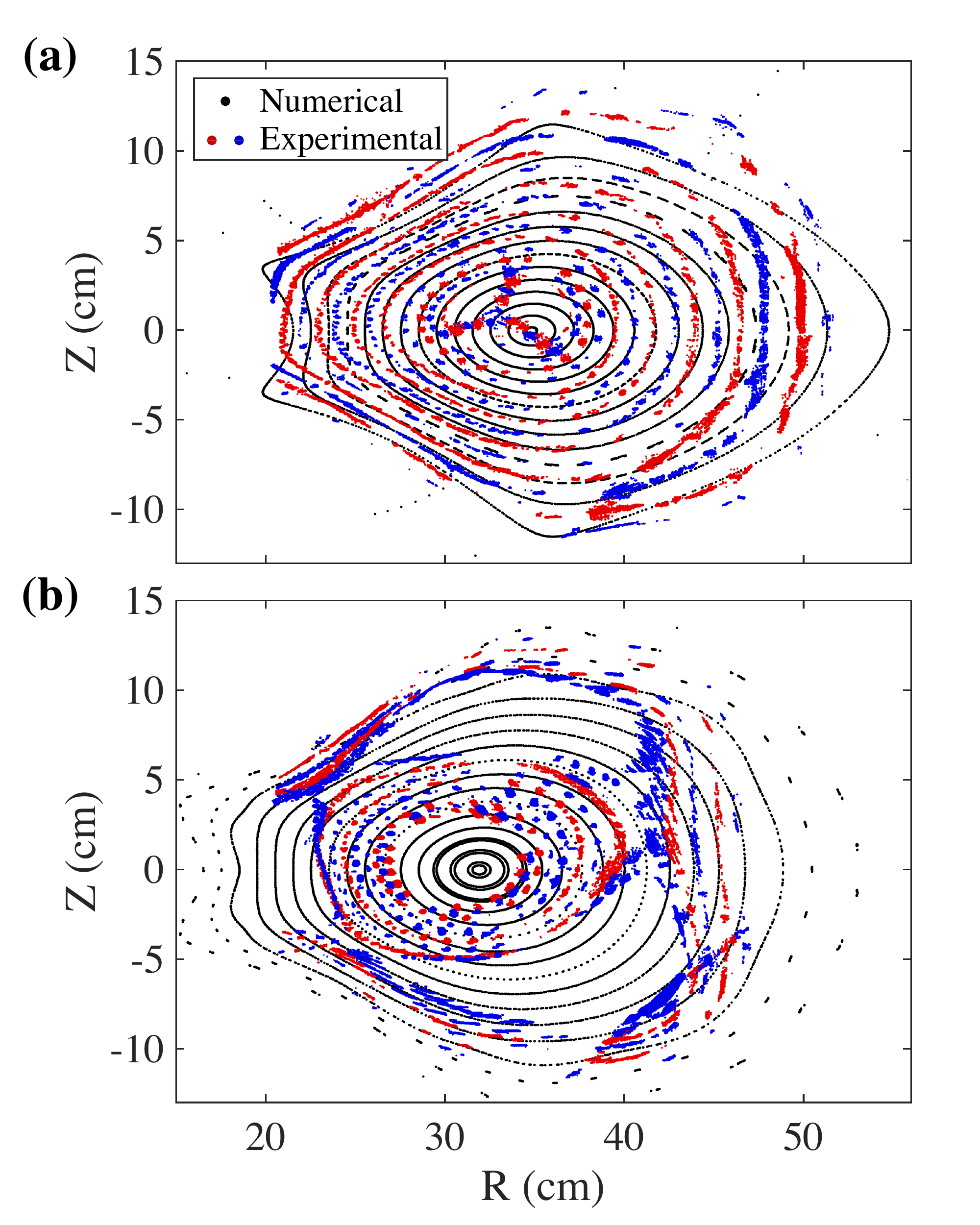}
    \caption{Comparison of numerical Poincar\'{e} plots (black dots) with 
             experimental results (adjacent surfaces are shown in alternating 
             red and blue for clarity). Numerical data were determined assuming
             that the coils were perfectly aligned.
             (a) $I_{IL}/I_{PF} = 3.68$; 
             (b) $I_{IL}/I_{PF} = 3.18$}
    \label{fig:ideal_poinc_comp}
    \end{center}
\end{figure}

Two examples of experimentally measured Poincar\'{e} cross-sections are shown 
in Fig.~\ref{fig:ideal_poinc_comp}, overlaid with numerically computed 
Poincar\'{e} data for the corresponding current-ratios. The numerical data,
determined using a Biot-Savart field line tracer described in 
Ref.~\cite{lazerson2016},
were generated assuming that the coils were perfectly aligned. 
Note that the qualitative agreement is poor: 
the outboard side of the surfaces at $I_{IL}/I_{PF} = 3.68$ 
(Fig.~\ref{fig:ideal_poinc_comp}a) are flatter and have more 
vertical elongation than the experimental surfaces.
In addition, 
the experimental data for $I_{IL}/I_{PF} = 3.18$ 
(Fig.~\ref{fig:ideal_poinc_comp}b) have prominent 
islands that are not predicted numerically. 

\begin{figure}
    \includegraphics[width=0.5\textwidth]{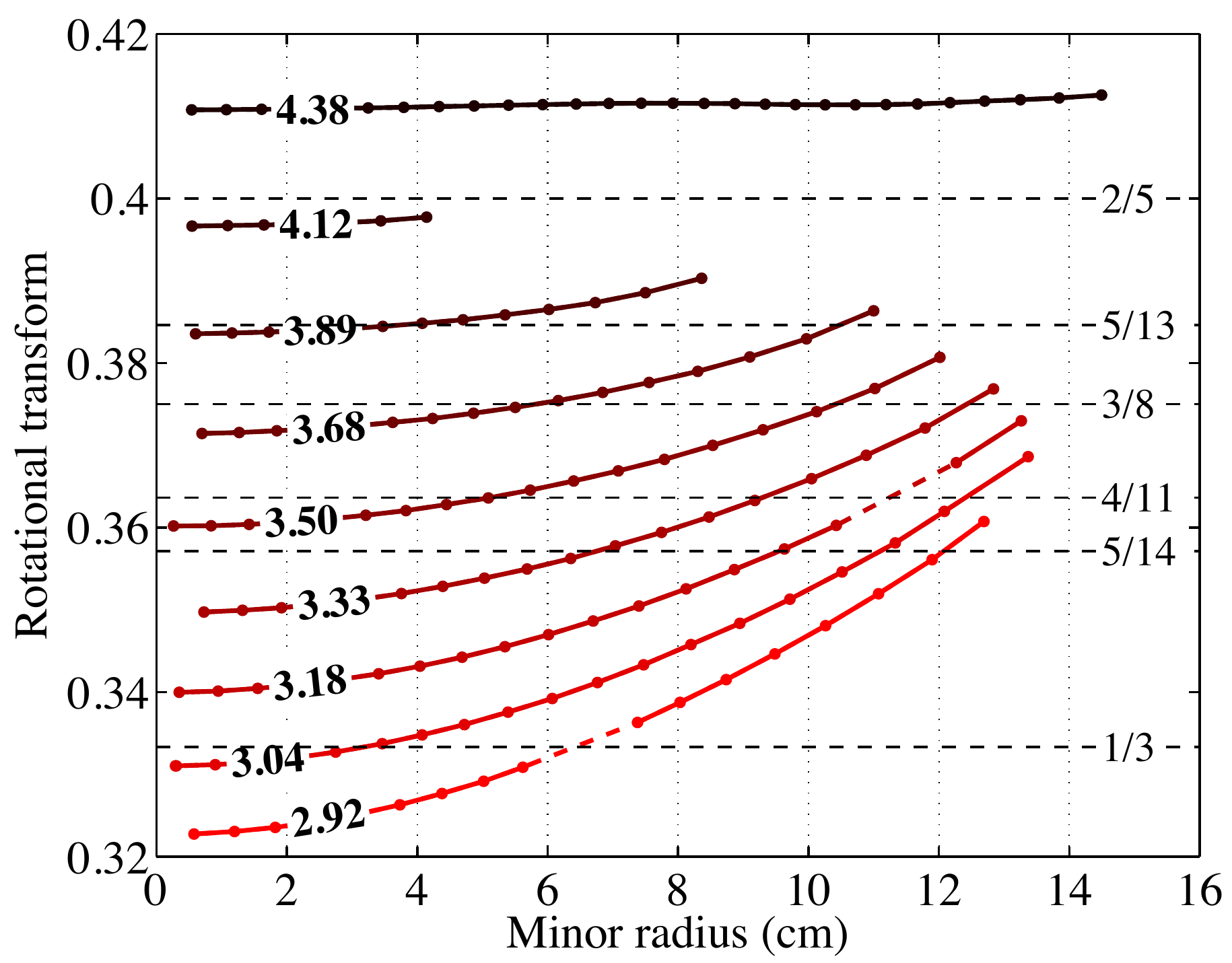}
    \caption{Plots of \sout{$\iota$} profiles for selected current-ratios 
             in CNT in the $78^{\circ}$ IL coil configuration. In the red
             curves, which are labeled with respective values of 
             $I_{IL}/I_{PF}$, each dot represents a closed flux surface and 
             dashed segments indicate island chains. Some low-order 
             rational numbers are shown as horizontal dashed lines. Note 
             that current-ratios of 3.18 and above are not predicted to contain 
             \sout{$\iota$}~=~1/3, contrary to measurements.}
    \label{fig:iota_unp}
\end{figure}

In addition to the geometric disagreements in the Poincar\'{e} sections at
toroidal angle $\phi = 90^\circ$, the measured rotational transform 
\sout{$\iota$} has been observed to differ from numerical predictions. 
While detailed profiles of \sout{$\iota$} have not been measured in CNT,
it is possible to identify low-order rational surfaces through 
field line visualizations \cite{brenner2008}. In this technique, the electron 
gun is operated while the vacuum vessel is back-filled to a neutral pressure 
between $10^{-5}$ and $10^{-4}$ Torr. At these pressures, electron-neutral 
collisions are frequent enough that the path of the electron beam emits a glow
that is visible to the naked eye. In the vicinity of a low-order rational 
surface, the beam can be seen to strike the back side of the electron gun 
after a finite number of toroidal transits. For example, for
$I_{IL}/I_{PF} < 3.5$, a region exists in which the electron beam is observed
to strike the electron gun after three toroidal transits, indicating the 
presence of a surface (or island chain) with \sout{$\iota$}~=~1/3. On the other
hand, numerically calculated \sout{$\iota$} profiles (Fig.~\ref{fig:iota_unp})
only contain \sout{$\iota$}~=~1/3 for $I_{IL}/I_{PF} < 3.18$.
Evidently, an EF is causing a systematic offset of the rotational transform.

The observed discrepancies between experiment and calculations motivated a 
more detailed study of the possible sources of EFs.

\section{Photogrammetry}
\label{sect:photogrammetry}

\begin{figure}
    \begin{center}
        \includegraphics[width=0.45\textwidth]{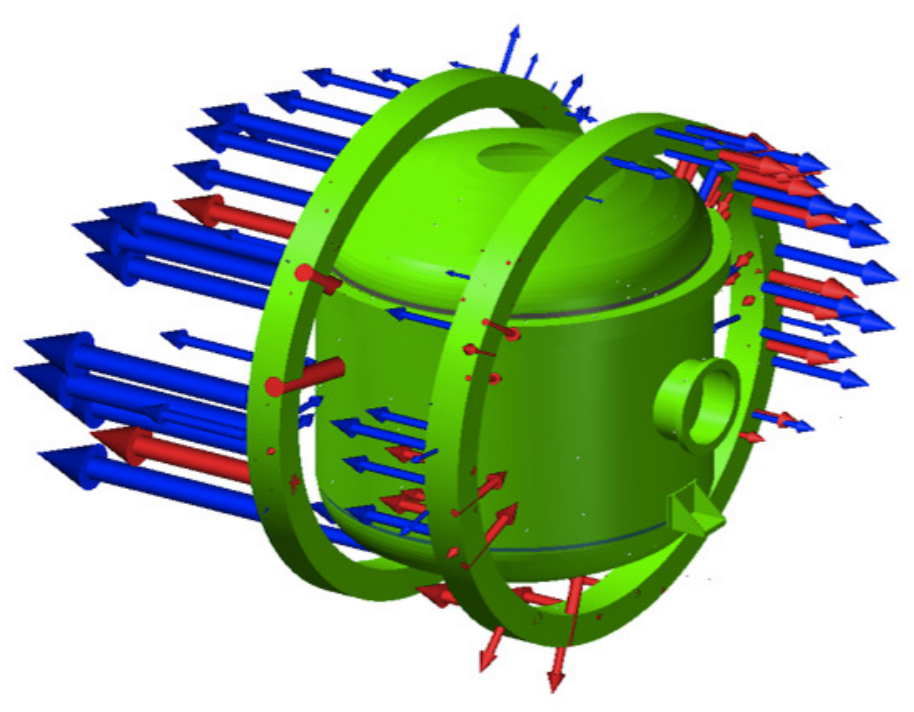}
        \caption{Rendering of the CNT vacuum vessel and PF coils overlaid with 
                 arrows showing the displacements of the coils from their 
                 design positions. Each arrow corresponds to one photogrammetric
                 marker. Arrow lengths are to scale with one another but are 
                 exaggerated relative to the vacuum vessel and coils. The 
                 largest arrow shown has a length of $44 \pm 3$ mm. Blue
                 (red) arrows represent deviations in the same (opposite) 
                 direction of the vector normal to the surface where they
                 originate.} 
        \label{fig:photogrammetry}
    \end{center}
\end{figure}

Displacements of the PF coils from their design positions were measured 
directly using photogrammetry. In this procedure, reflective markers were 
affixed to all parts of the PF coils as well as the exterior of the vacuum 
vessel. The vessel and coils were then photographed from several different 
angles. The photographs were analyzed by V-STARS software (produced by 
Geodetic Systems, Inc.), which generated a point cloud giving the 
relative positions of the markers in 3D space. The point cloud was then 
fitted to the nominal surfaces of the machine as provided by a CAD file.
The result was a set of displacement vectors (Fig.~\ref{fig:photogrammetry})
for each marker corresponding to the displacement of its respective part of 
the coil from the design location. Displacement vectors that deviated 
significantly from those of neighboring markers were viewed as erroneous and 
were ignored in subsequent analysis.

An attempt was also made to measure the positions of the IL coils using a
Romer six-axis measuring arm. However, limited access to the IL coils 
restricted their measurements to a small region of each coil. The measurements
obtained were sufficient to estimate the location of the center 
of the device to within less than 1 mm (which was then used as a reference for 
the PF coil marker offsets), but did not yield accurate estimates of 
the misalignments of the IL coils relative to one another. We will note that it
is possible to configure the chamber to permit full access for the Romer arm,
but this could not be done on the time scale of the arm's availability for the 
measurements described in this paper and will be an objective for future work.

According to the marker displacements determined by the software, certain 
regions of the PF coils are separated by as much as $44 \pm 3$ mm from their 
intended locations, which is well outside of the 10 mm tolerance specified for 
construction. It is noteworthy that CNT's coils still produce good flux 
surfaces despite having displacements of this magnitude.

The observed misalignments of the PF coils are believed to have arisen during 
a procedure to change the IL coil configuration from 
$\theta_{tilt} = 64^\circ$ to $\theta_{tilt} = 78^\circ$. 
The PF coils must be tilted out of the way to permit access to the IL coils, 
and they sit on hinged support structures for this purpose. By the end of the
procedure, the structures had visibly deformed due to internal stresses 
arising from the tilting, likely affecting the positions of the PF coils 
themselves.

To determine the effects of the measured PF coil displacements on the flux 
surfaces, new field line traces were conducted with the PF coils offset 
according to the photogrammetry data. For these calculations, the PF coils 
were still assumed to be circular and planar but were tilted and translated 
so as to best fit the displacement vectors. According to these fits, the 
northern ($z>0$) and southern ($z<0$) PF coils were translated 5 mm
and 22 mm respectively, and their axes were tilted by $1.0^\circ$ and 
$1.3^\circ$ respectively.

\begin{figure}
    \includegraphics[width=0.5\textwidth]{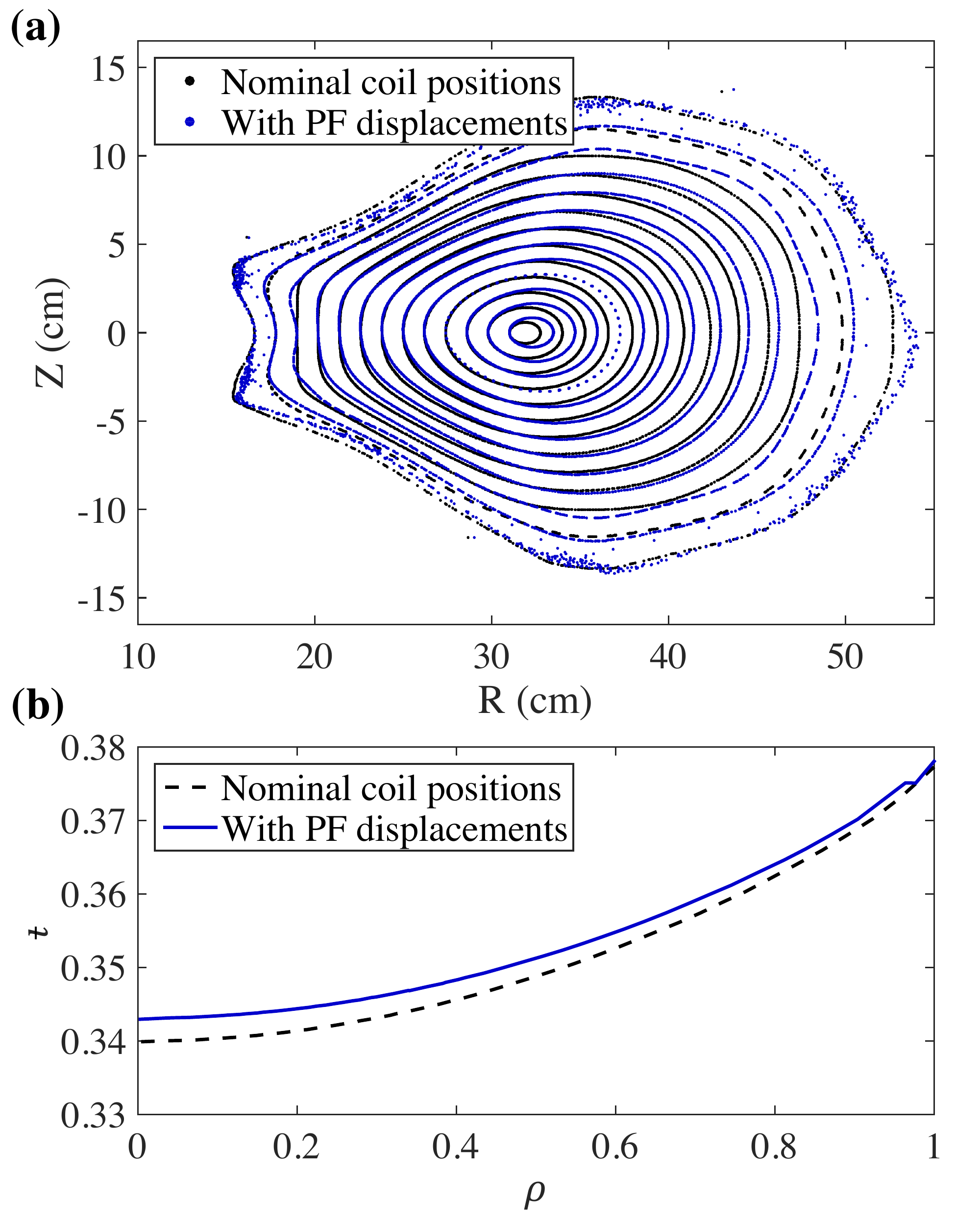}
    \caption{(a) Numerically generated Poincar\'{e} plots at the 
             $\phi = 90^\circ$ cross-section with $I_{IL}/I_{PF} = 3.18$ 
             from a configuration incorporating the measured PF coil 
             displacements (blue) and from the design configuration (black). 
             (b) Calculated profiles of \sout{$\iota$} for this configuration.}
    \label{fig:pfc_disp_compare}
\end{figure}

A comparison of the numerical Poincar\'{e} plots with and without the 
displacements is shown in Fig.~\ref{fig:pfc_disp_compare} using 
$I_{IL}/I_{PF} = 3.18$. Although the plots are not identical (in 
particular, the magnetic axis has shifted slightly outboard and field lines 
initiated at the same locations do not overlap perfectly),
the effects of the PF coil displacements do 
not appear to be sufficient to explain the observed experimental discrepancies.
In particular, the slight increase in the \sout{$\iota$} profile is actually the
opposite effect of what would be needed to explain the \sout{$\iota$} = 1/3 
island chain seen at an intermediate minor radius in 
Fig.~\ref{fig:ideal_poinc_comp}b.

Since the PF coil displacements were not enough to explain the observed 
differences in Poincar\'{e} cross-sections, additional sources of field error 
were sought.

\section{Numerical study of \sout{$\iota$} sensitivity}
\label{sect:numerical_studies}

In the search for additional sources of field error, the most likely candidate 
was thought to be displacements to the IL coils. Previous work 
\cite{kremer_thesis} has shown that a translational displacement of either of
the IL coils has a significantly greater effect on the magnetic geometry than 
the same displacement applied to either of
the PF coils. (For this reason, the tolerance for the IL coils was set at 2 mm,
one-fifth of the PF coil tolerance.) 

\begin{figure}
    \includegraphics[width=0.5\textwidth]{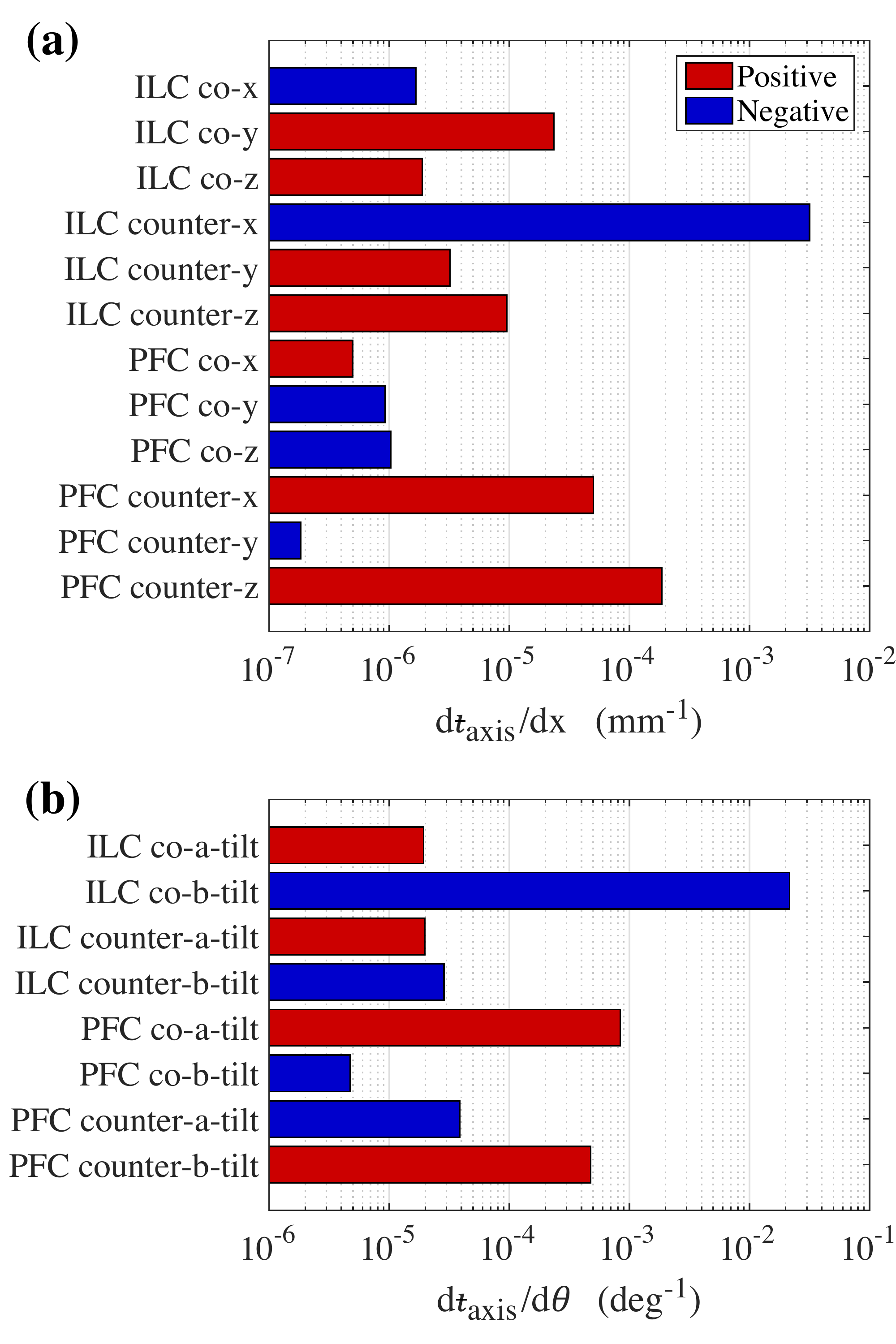}
    \caption{Derivatives of \sout{$\iota$}$_{axis}$ with respect
             to the magnitude of different classes of coil displacements. 
             (a) Translations of the coils in the Cartesian directions 
             illustrated in Fig.~\ref{fig:coil_schematic}.
             (b) Tilts of the coils in the directions of their respective
             $\hat{a}$ and $\hat{b}$ axes. The $\hat{a}$ and $\hat{b}$ axes
             for the IL coils are illustrated in Fig.~\ref{fig:coil_axes}; 
             for both PF coils, $\hat{a}=\hat{x}$ and $\hat{b}=\hat{y}$.
             A more detailed description of each displacement class is given
             in the text.}
    \label{fig:diota_compare}
\end{figure}

To identify the possible causes of the systematic offset in \sout{$\iota$} 
discussed in Section \ref{sect:flux_surf_meas}, a series of numerical field 
line traces was computed for different classes of IL coil displacements. Field 
lines were also traced for similar classes of PF coil displacements for 
comparison. The classes of displacements were sorted into symmetric (equal coil 
movement in opposite directions) and antisymmetric (equal coil movement in the 
same direction). For each class of displacement, \sout{$\iota$} profiles were 
calculated for a series of magnitudes of the displacement. To represent the 
calculated trends in \sout{$\iota$}, Fig.~\ref{fig:diota_compare} shows 
derivatives of \sout{$\iota$}$_{axis}$ (\textit{i.e.}, the limit of 
\sout{$\iota$} as the minor radius approaches zero) with respect to each type 
of displacement. Derivatives were estimated by second-order finite differences 
about the nominal positions of the coils at $I_{IL}/I_{PF} = 3.68$.

The displacements plotted in Fig.~\ref{fig:diota_compare} collectively represent
the twenty degrees of freedom that the four coils have for rigid motion (five 
for each coil; rotation of a coil about its axis is ignored due to symmetry).
The translational displacements shown in Fig.~\ref{fig:diota_compare}a have 
fairly intuitive interpretations. A positive ``ILC co-$x$'' displacement, for 
example, involves motion of both the IL coils in the positive $x$ direction as 
indicated in Fig.~\ref{fig:coil_schematic}. An ``ILC counter-$x$'' displacement,
on the other hand, involves motion in opposite ways in the $x$ direction, with
positive displacement indicating the IL coils moving apart and negative 
displacement indicating the IL coils moving toward one another. 

\begin{figure}
    \begin{center}
    \includegraphics[width=0.3\textwidth]{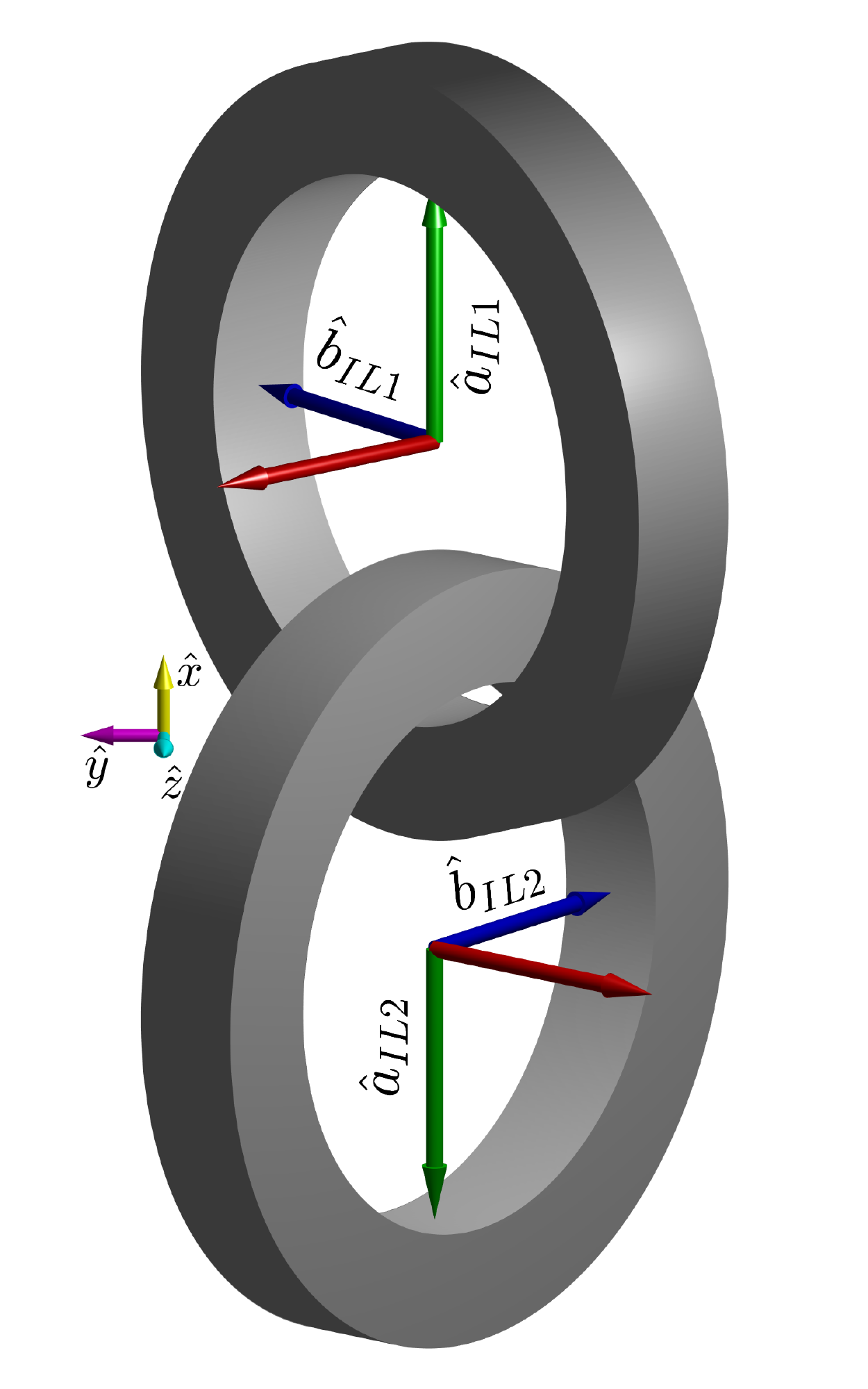}
    \caption{Schematic of the two IL coils along with their respective 
             axes of symmetry (red arrows), $\hat{a}$ vectors (green arrows), 
             and $\hat{b}$ vectors (blue arrows) as described in the text. 
             The $\hat{x}$, $\hat{y}$, and $\hat{z}$ directions as defined in 
             Fig.~\ref{fig:coil_schematic} are shown in yellow, magenta, and 
             cyan respectively.}
    \label{fig:coil_axes}
    \end{center}
\end{figure}

The angular displacements shown in Fig.~\ref{fig:diota_compare}b represent 
tilts of the coil in two orthogonal directions denoted by $\hat{a}$ and 
$\hat{b}$. The $\hat{a}$ and $\hat{b}$ vectors for the IL coils are illustrated 
in Fig.~\ref{fig:coil_axes}. ``ILC co-b-tilt'' displacement, for example, 
refers to both IL coils tilting by some angle $\theta$ such that their axes 
rotate toward their 
respective $\hat{b}$ unit vectors. A change in ILC co-b-tilt is equivalent to
an adjustment of $\theta_{tilt}$ (nominally $78^\circ$). 
``ILC counter-a-tilt'' displacement, as another example, refers to the IL 
coils tilting such that IL1 tilts its axis in the positive $\hat{a}_{IL1}$ 
direction and IL2 tilts its axis in the negative $\hat{a}_{IL2}$ direction.
For both PF coils, $\hat{a}=\hat{x}$ and $\hat{b}=\hat{y}$.

From Fig.~\ref{fig:diota_compare}, it is evident that the two classes of
displacements that have the greatest influence on \sout{$\iota$} are the 
separation of the IL coils (along the $x$ axis) and the tilt angle between 
the IL coils. Another less-prominent influence arises from 
the separation of the PF coils. This is roughly equivalent to changing the 
PF coil current, as it strengthens or weakens the Helmholtz field created 
by the coils, and illustrates how adjusting the PF current relative to the IL 
current serves as a fine adjustment to the iota profile.

The relative sensitivity of \sout{$\iota$} to the different classes 
of coil displacements in Fig.~\ref{fig:diota_compare} suggests that, if the 
offset of the \sout{$\iota$} profile is indeed caused by coil displacements, 
it is very likely that displacements of the IL coils play a significant role.
This observation motivated a more detailed study of IL coil displacements to 
be described in the following section.

\section{Optimization}
\label{sect:optimization}

\subsection{Optimization procedure}
\label{subsect:opt_proc}

The optimization method used in this study tweaks the coil positions 
in small increments until the calculated Poincar\'{e} 
data resulting from the perturbation matches with the experimentally obtained 
Poincar\'{e} cross-sections. The procedure is conceptually similar to methods 
used in some plasma equilibrium reconstruction codes such as EFIT and V3FIT 
\cite{lao1985,hanson2009}, which determine plasma equilibrium parameters that 
fit to diagnostic signals. In this procedure, Poincar\'{e} cross-sections take
the role of the diagnostic signals, whereas coil displacements have the role 
of the equilibrium parameters. 

For the optimizations described in this paper, the observed Poincar\'{e} 
cross-sections are mapped into a vector $\mathbf{X}$ of discrete geometric 
parameters, which may be thought of equivalently as physics parameters derived 
from flux surface topology. The definitions of these parameters are provided in 
\ref{sect:parametrization}, and the procedure for determining 
them from experimental and numerical Poincar\'{e} data is given in 
\ref{sect:coeff_determination}. The coil misalignments were encapsulated in 
a vector $\mathbf{p}$ of displacement parameters (\textit{i.e.}, engineering 
parameters). The work in this paper used two variants of $\mathbf{p}$, both 
of which will be described later in this section.

An optimization begins with target geometric parameters $\mathbf{X^*}$ for 
the desired Poincar\'{e} cross-section geometry and an initial guess 
$\mathbf{p}_0$ of coil displacement parameters. The geometric parameters
associated with $\mathbf{p}_0$ (or any set of dispacement paramameters
$\mathbf{p}$) are then $\mathbf{X}(\mathbf{p})$. The discrepancy 
$\mathbf{F}(\mathbf{p})$ between $\mathbf{X^*}$ and 
$\mathbf{X}(\mathbf{p})$ is defined as

\begin{equation}
    F_i(\mathbf{p}) = X_i(\mathbf{p}) - X_i^*
\label{eqn:fvec}
\end{equation}

The optimization algorithm attempts to find $\mathbf{p}^*$ such that 
$\mathbf{F}=0$ using the Newton-Raphson method \cite{press1992}. Expanding
$\mathbf{F}$ about a coil configuration $\mathbf{p}$ that differs from 
$\mathbf{p}^*$ by $\delta\mathbf{p}$, i.e.,

\begin{equation}
    \mathbf{F}(\mathbf{p}+\delta\mathbf{p}) = 0 = \mathbf{F}(\mathbf{p}) 
        + J\delta\mathbf{p} + O\left(\delta\mathbf{p}^2\right),
\end{equation}

\noindent the iterative Newton step $\delta\mathbf{p}$ is given by a
solution to the equation

\begin{equation}
     \mathbf{F}(\mathbf{p}) = -J\delta\mathbf{p}.
\end{equation}

\noindent Here, $J$ is the Jacobian,

\begin{equation}
    J_{ij} = \frac{\partial F_i}{\partial p_j} 
\label{eq:jacobian}
\end{equation}

In general, $\mathbf{p}$ is not of the same dimension as $\mathbf{F}$. We 
thus find the optimal $\delta\mathbf{p}$ through linear least-squares. 
The Best Linear Unbiased Estimator (BLUE) for $\delta\mathbf{p}$ is that which 
minimizes $\chi^2$ \cite{jones2006}:

\begin{equation}
    \chi^2 = \mathbf{F}^T C^{-1} \mathbf{F}
\label{eqn:chi2}
\end{equation}

\noindent where $C_{ij} = cov\left(X_i^*,X_j^*\right)$ is the covariance 
matrix for the target parameters $\mathbf{X^*}$. This estimator for 
$\delta\mathbf{p}$ is given by

\begin{equation}
    \delta\mathbf{p} = -\left(J^TC^{-1}J\right)^{-1}J^TC^{-1}\mathbf{F}.
\label{eqn:blue}
\end{equation}

\subsection{Numerical considerations}
\label{subsect:numerical}

\subsubsection{The covariance matrix}

For experimental Poincar\'{e} data, the covariance matrix $C$ is estimated by 
evaluating $\mathbf{X}$ for multiple samples of the pixels obtained from 
composite flux surface images. For each sample, pixels are selected randomly 
with replacement (bootstrapping) \cite{islr2013}. The number of pixels 
available to sample from a particular flux 
surface in the images used for this analysis ranged from 300 to more than 
10,000. Each sample consists of 400 pixels per flux surface.

For the verification studies conducted in Sec.~\ref{subsect:verification}, 
the optimizer was programmed to fit the coils to numerically determined 
Poincar\'{e} data (\textit{i.e.}, a manufactured solution). To obtain a 
covariance matrix for numerical Poincar\'{e} data, multiple Poincar\'{e} plots
are obtained by tracing field lines from randomized initialization points. 
Each sample consists of a trace of 200 field lines followed for 200 toroidal 
revolutions. (A certain percentage of the field lines will terminate if they 
are initialized outside of the last closed flux surface; these are ignored.)
 
As indicated in Eq.~\ref{eqn:blue}, $C$ must be inverted for the analysis. This 
can introduce significant numerical error if $C$ is ill-conditioned. To 
mitigate this risk, rather than directly inverting $C$, its inverse is 
approximated through singular value decomposition: 

\begin{equation}
    C = U\Sigma U^T
\end{equation}

\noindent Here, the orthogonal matrices $U$ to the left and right of $\Sigma$ 
are the same due to the inherent symmetry of $C$. $\Sigma$ is a diagonal 
matrix of the singular values. If $C$ is ill-conditioned, the lowest 
singular value(s) will be much less than the greatest singular value. 
As a workaround, 
a cutoff is imposed such that all singular values of $C$ less than a 
factor $1/\kappa_{co}$ of the greatest singular value are set to zero.
The pseudoinverse of this approximation is given by

\begin{equation}
    C^{-1} \approx U\Sigma^{-1} U^T,
\label{eqn:Cinv}
\end{equation}

\noindent where $\Sigma^{-1}$ is a diagonal matrix in which the $n^{th}$ 
nonzero diagonal element is equal to the inverse of the $n^{th}$ nonzero 
diagonal element of $\Sigma$. For the work described in this paper, 
$\kappa_{co}$ was chosen empirically to be $10^8$ for the 
calculations in this paper as the lowest value for which the Newton direction 
was not noticeably compromised.

\subsubsection{The Jacobian}

The elements of the Jacobian in this implementation are computed by 
second-order finite differences:

\begin{equation}
    J_{ij} = \frac{F_i\left(\mathbf{p}+\Delta\mathbf{p}_j\right)
                      - F_i\left(\mathbf{p}-\Delta\mathbf{p}_j\right)}
                  {\left|\Delta\mathbf{p}_j\right|}
\end{equation}

Note that, due to its dependence on $\mathbf{X^*}$ via $\mathbf{F}$, $J_{ij}$ 
contains random error and may be correlated with other matrix elements. 
Because of this, Eq.~\ref{eqn:blue} does not, strictly speaking, give the 
BLUE for $\delta\mathbf{p}$, since one of the underlying assumptions is that 
$J$ is non-random. Nevertheless, we have found that this estimator is 
sufficient in many cases as long as the finite differencing interval 
$\Delta\mathbf{p}$ is sufficiently large. 

\subsubsection{Line-search}

If $\mathbf{F}$ depends nonlinearly on $\mathbf{p}$, there is a risk that the
Newton step $\delta\mathbf{p}$ will overshoot a local minimum in $\chi^2$. 
To rectify such occurrences, a line-search algorithm checks the 
Newton step after each iteration to ensure that the average rate of decrease 
in $\chi^2$ over the interval $\delta\mathbf{p}$ is at least as great as the 
gradient of $\chi^2$ evaluated at the starting point $\mathbf{p}$ of the 
iteration. If this condition is not met, the algorithm samples $\chi^2$ at 
shorter distances along the direction of the Newton step to identify the 
local minimum. More details on this algorithm are given in 
Ref.~\cite{press1992}.

\subsection{Verification}
\label{subsect:verification}

To verify the performance of the algorithm, some tests were conducted in which
the code was used to solve for a known coil displacement $\mathbf{p}^*$ 
using target parameters $X^*$ generated from a field-line trace that used
$\mathbf{p}^*$ as its input. 

One early test was meant to determine whether the algorithm could distinguish 
two classes of displacements that produced qualitatively similar outcomes. 
One such pair of displacement classes is (1) separation of the coils along the 
$x$ axis and (2) adjustment of $\theta_{tilt}$. Incidentally, these two classes
were also found to have the greatest impact on rotational transform 
(Fig.~\ref{fig:diota_compare}), a strong determiner of the cross-section 
geometry. A $\mathbf{p}$ vector was thus defined with just two components:

\renewcommand{\arraystretch}{1.5}
\begin{equation}
\mathbf{p} = \left[ \begin{array}{c c} 
              \frac{1}{2c_x}\left(\Delta x_{IL1} - \Delta x_{IL2}\right) \\
              \frac{1}{c_\theta}\Delta\theta_{tilt} \end{array} \right]
\end{equation}

\noindent Here, $\Delta x_{IL1}$ and $\Delta x_{IL2}$ are the displacements
along the machine $x$ axis of the first and second IL coil, respectively, from
their nominal positions. $\Delta\theta$ is the displacement of the coil angle 
from its nominal value of $78^\circ$. $c_x$ and $c_\theta$ are scaling constants
equal to 1 m and 1 radian, respectively. The target parameters $\textbf{X}^*$
were generated using $p_1 = 0$ and $p_2 = 8\times10^{-3}$; \textit{i.e.}, 
no discrepancy in coil separation and a decrease of $0.46^\circ$ in 
$\theta_{tilt}$. The initial guess $\mathbf{p}_0$ was (0, 0) and the finite 
differencing interval for both components was $10^{-3}$.

\begin{figure}
    \includegraphics[width=0.5\textwidth]{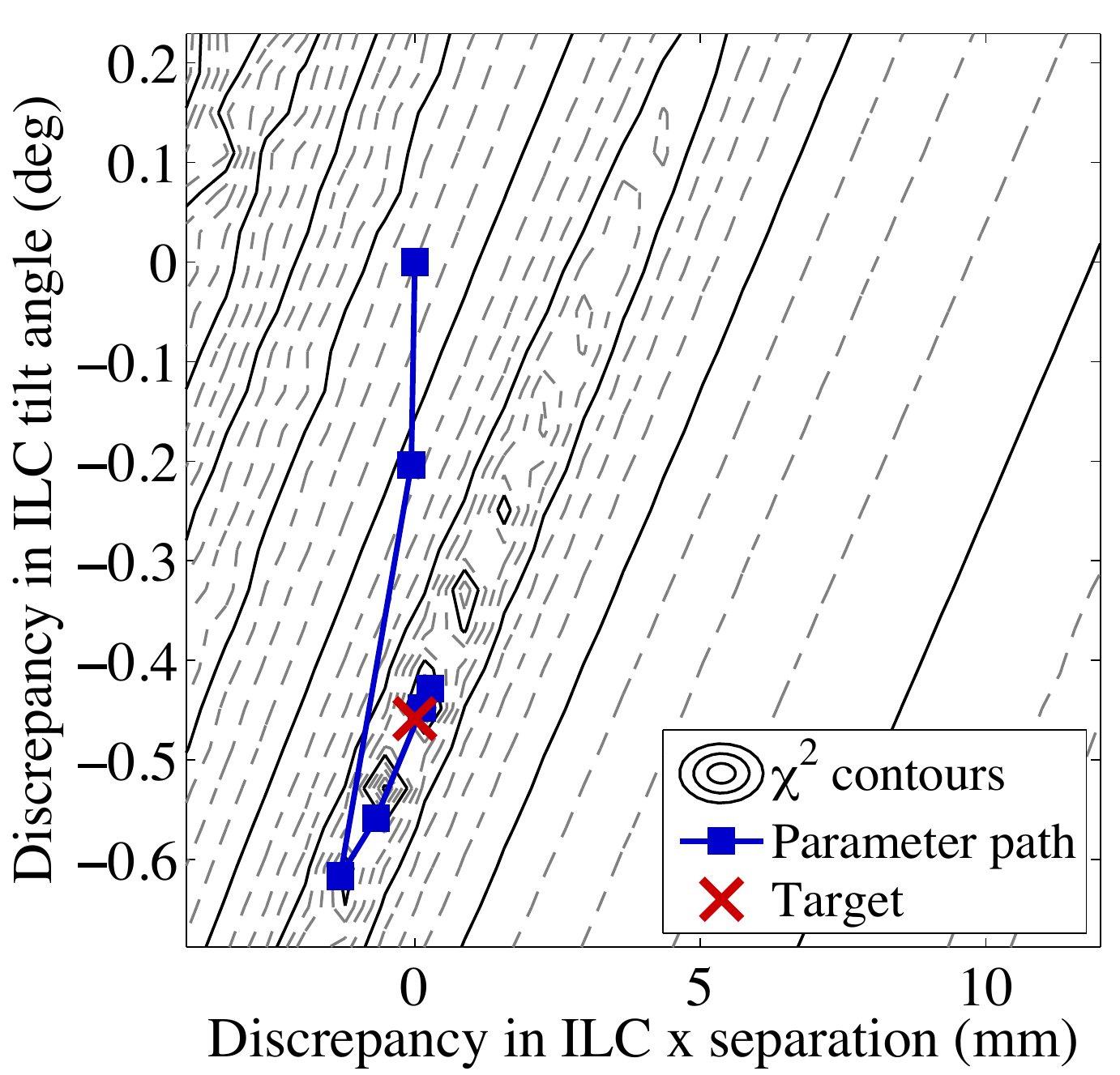}
    \caption{Contour plot of $\chi^2$ and the path taken by the optimizer
             in a verification test with two free parameters. The initial guess
             $\mathbf{p}_0$ is (0, 0) and the target parameters are denoted
             by the red $\times$. Contours are on a logarithmic scale with 
             solid contours representing powers of 10.}
    \label{fig:contours_tilt_dx}
\end{figure}

The results of this test are shown in Fig.~\ref{fig:contours_tilt_dx}. 
Note that the contours of $\chi^2$ exhibit a shallow valley
along a line consisting of a family of displacements that would 
result in similar Poincar\'{e} geometry. Nevertheless, the algorithm
succeeded in moving the components of $\mathbf{p}$ to within 
one finite differencing interval of the target values on the fourth 
iteration.

\begin{figure}
    \includegraphics[width=0.5\textwidth]{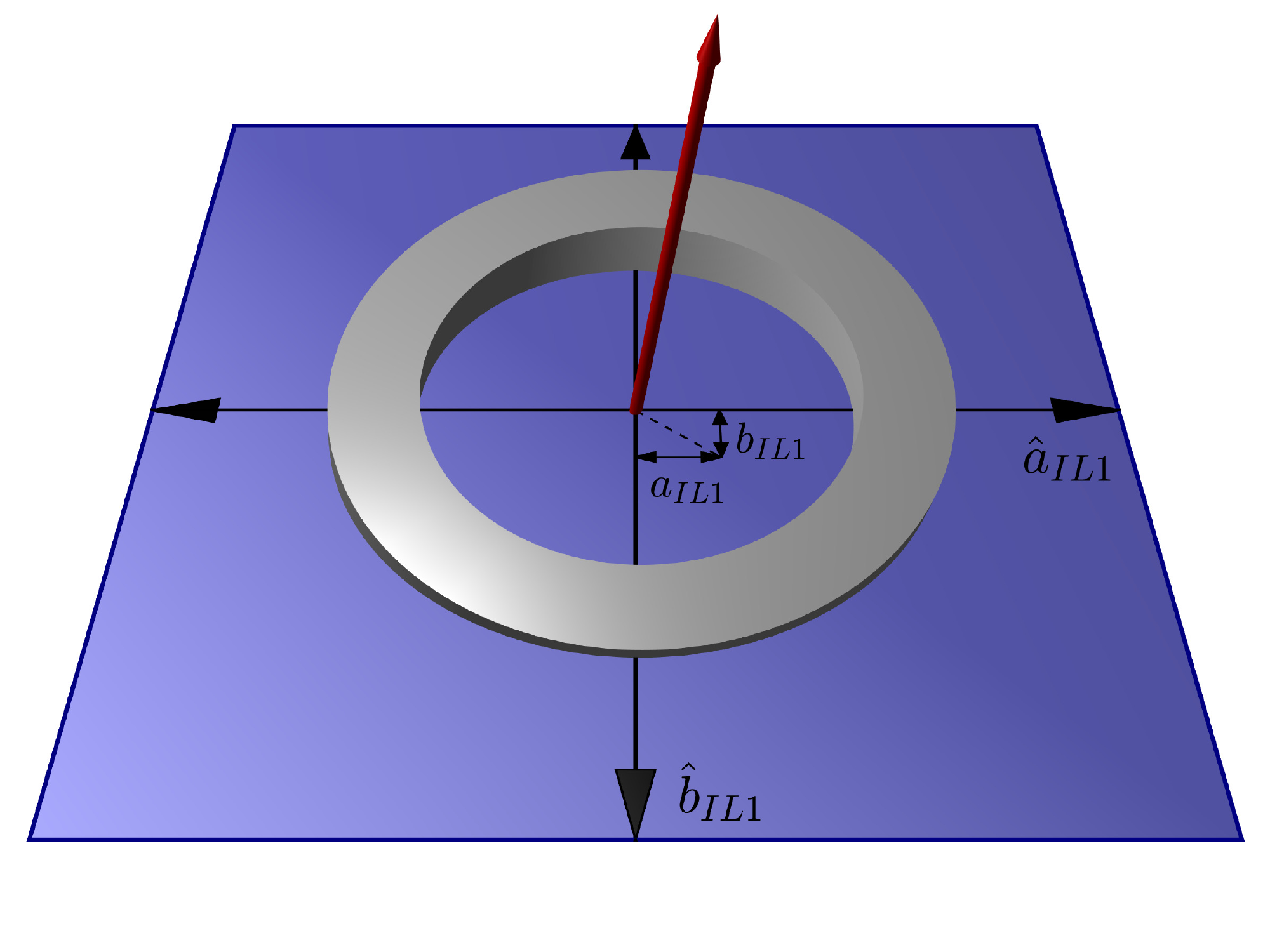}
    \caption{Schematic illustrating the definitions of the $a$ and 
             $b$ components of the angular displacement parameters 
             of the coils. In this image, the IL1 coil is used as an example.
             The blue surface represents the plane of the unperturbed coil, 
             and the red arrow represents a unit vector along the axis of 
             the perturbed coil.}
    \label{fig:coil_disp}
\end{figure}

Subsequent verifications tested the ability of the code to identify similar 
coil displacements when more coil parameters were free. In these tests, 
the IL coils were permitted to undergo rigid rotations and transformations; 
hence, each coil was allowed five degrees of freedom: three translational 
and two angular. The translational parameters, 
$x$, $y$, and $z$, are simply Cartesian displacements along 
the respective unit vectors indicated in Fig.~\ref{fig:coil_schematic}.
The angular parameters are illustrated in Fig.~\ref{fig:coil_disp}. As shown 
in the diagram, $a_{IL}$ and $b_{IL}$ are 
orthogonal projections of a unit vector representing the 
\textit{perturbed} coil axis onto the plane of the \textit{unperturbed} coil. 
The $a_{IL1}$ and $b_{IL1}$ components correspond to the directions 
$\hat{a}_{IL1}$ and $\hat{b}_{IL1}$ illustrated in Fig.~\ref{fig:coil_axes}.
The polar displacement angle of the perturbed axis of the first IL coil 
can thus be computed as 
$\arcsin{\left(\sqrt{{a_{IL1}}^2 + {b_{IL1}}^2}\right)}$. 

\begin{figure}
    \includegraphics[width=0.5\textwidth]{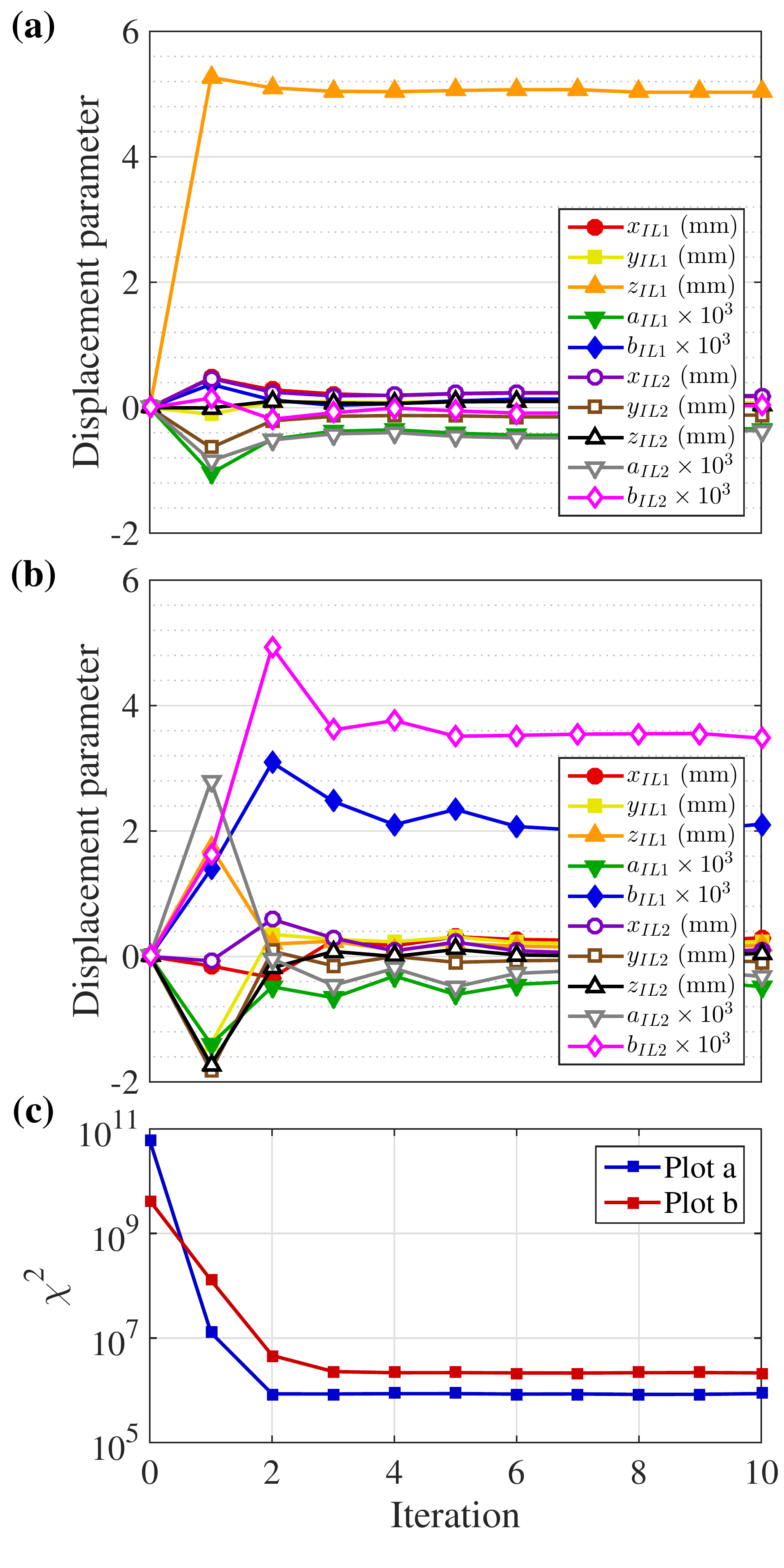}
    \caption{Results of optimizations in which the interlocked coils were 
             free to move in 10 parameters to match Poincar\'{e} data 
             generated from made-up coil perturbations. (a) Parameter evolution 
             toward targets of $z_{IL1}=5$ mm and the rest of the 
             parameters zero. (b) Parameter evolution toward targets
             of $b_{IL1} = 0.002$, $b_{IL2} = 0.004$, and the rest zero. 
             (c) Descent of $\chi^2$ for both optimizations.}
    \label{fig:verification}
\end{figure}

Fig.~\ref{fig:verification} shows the outcomes of two calculations that 
optimized the ten IL coil parameters as described above to a target 
$\mathbf{X^*}$ vector generated numerically from a chosen set $\mathbf{p}^*$
of displacements. In Fig.~\ref{fig:verification}a, the only nonzero component
of $\mathbf{p}^*$ was $z_{IL1}$, which was set to 5 mm, a simple translation 
of the first IL coil. 
In Fig.~\ref{fig:verification}b, there were two nonzero components of 
$\mathbf{p}^*$: $b_{IL1}$ = 0.002 and $b_{IL2}$ = 0.004; \textit{i.e.}, 
both coils were tilted toward their respective $\hat{b}$ vectors, but by 
different amounts.

Both optimizations descended by multiple orders of magnitude in $\chi^2$ 
in the first three steps (Fig.~\ref{fig:verification}c), but 
afterward, $\chi^2$  flattened out: the 
optimizations did not converge any further toward $\mathbf{p}^*$. One 
possible explanation is that a relatively long finite differencing interval
$\Delta\mathbf{p}$ was used to determine the Jacobian in each step: all 
translational 
($x$, $y$, and $z$) components used an interval of 1 mm, and the angular 
components ($a$ and $b$) used an interval of 0.001. All of the final values
$p_i$ of the components are well within $\Delta p_i$ of the target 
values $p_i^*$ (Fig.~\ref{fig:verification}a-b). Another contributing
factor may be the numerical uncertainty of $\mathbf{F}$.

\subsection{Inferring coil displacements from experimental data}
\label{subsect:opt_exp}

After the verification tests, the algorithm was applied to experimental 
Poincar\'{e} data. The target geometric parameters $\mathbf{X}^*$ were
determined from the cross-section data for the current-ratio 3.68 
(Fig.~\ref{fig:data2params}). This dataset was chosen over datasets from
other current-ratios for its abundance of
fully-characterized flux surfaces and for its lack of magnetic islands, two
attributes that facilitate accurate calculations of $\mathbf{X}^*$
(Sec.~\ref{subsect:islands}).
The PF coils were held fixed with displacements 
determined by a best-fit to the photogrammetry data discussed in 
Sec.~\ref{sect:photogrammetry}.
All ten parameters associated with the IL coils were free.
The initial guess $\mathbf{p}_0$ was the nominal IL coil configuration. 
As in the 10-parameter verifications conducted in 
Sec.~\ref{subsect:verification}, the finite differencing interval was 1 mm
for all translational motion and 0.001 for all angular displacements.
The set of displacements obtained
in the final iteration will be referred to hereafter as $\mathbf{p}^*_{3.68}$.

\begin{figure}
    \includegraphics[width=0.5\textwidth]{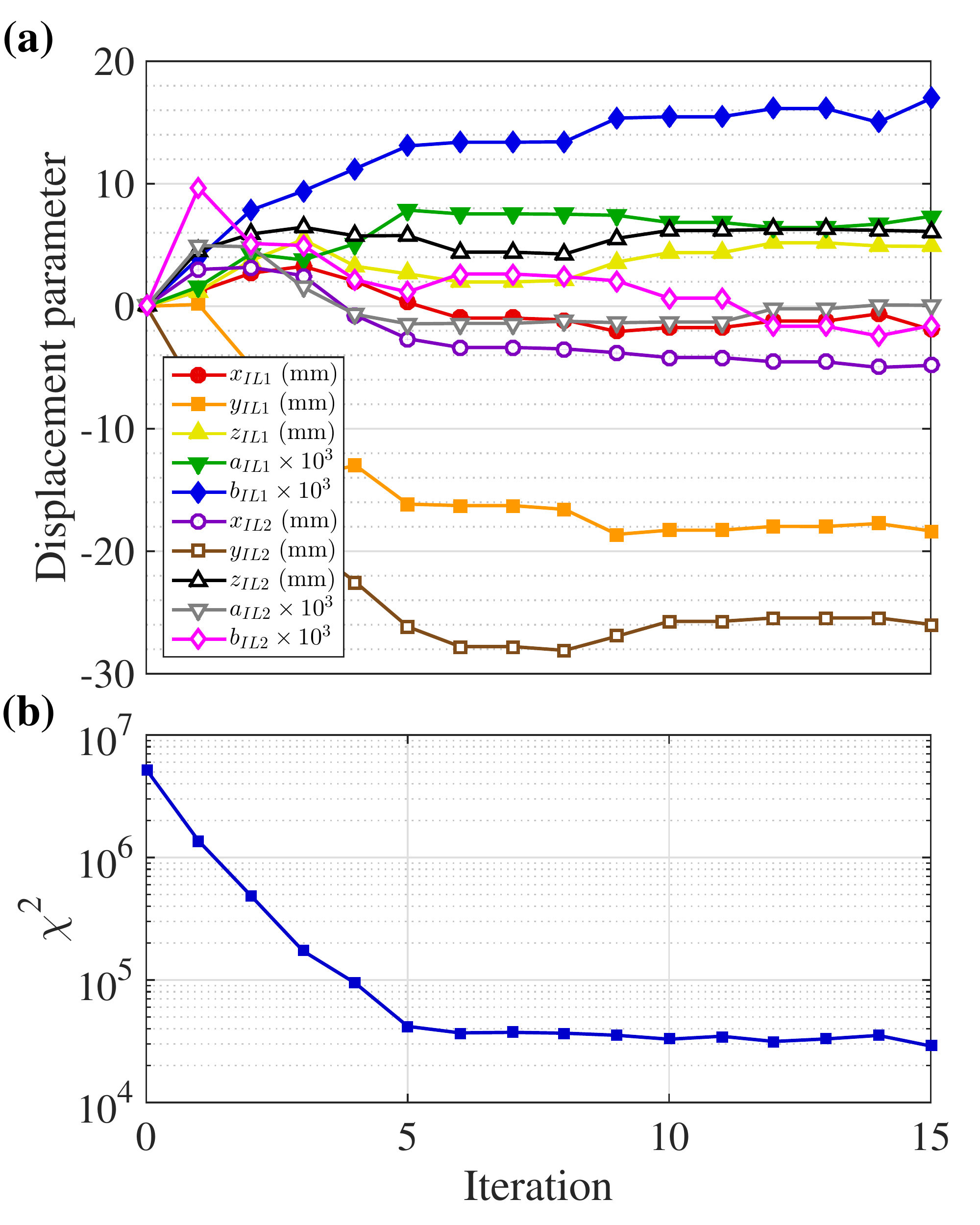}
    \caption{(a) Evolution of the displacement parameters of the IL coils 
             during an optimization using experimental Poincar\'{e} data from
             the 3.68 current-ratio. (b) Descent of $\chi^2$ associated 
             with the Poincar\'{e} data resulting from the displacements.}
    \label{fig:params_exp_10free}
\end{figure}

The shifts in IL coil positions during the fifteen steps of the optimization are
shown in Fig.~\ref{fig:params_exp_10free}a, with the accompanying descent of
$\chi^2$ shown in Fig.~\ref{fig:params_exp_10free}b. The largest translational moves from the starting positions
occurred in the negative $\hat{y}$ direction for both coils, averaging to 
-22 mm. The $\hat{z}$ displacement was nearly identical at around 6 mm for 
both coils, and the $\hat{x}$ displacement averaged to about 3 mm with a 3
mm ``counter-'' component (cf. the displacement classes in 
Fig.~\ref{fig:diota_compare}). It should be noted that, while the ``co-'' 
translational motion of the coils may indicate that the IL coils are indeed
off of their nominal positions, it may also reflect misalignment of the 
fluorescent rod.

The largest angular shift was of the first IL coil, which tilted about 
$1^\circ$ away from its nominal axis along $\hat{a}_{IL1}$, effectively 
decreasing $\theta_{tilt}$ to $77^\circ$. With the results of 
Fig.~\ref{fig:diota_compare} in mind, this shift, combined with the 
3 mm of counter-motion in the $\hat{x}$ direction, is likely responsible for 
most of the downward offset in \sout{$\iota$} observed experimentally.

During the optimization, the value of $\chi^2$ 
(Fig.~\ref{fig:params_exp_10free}b) for the Poincar\'{e} data 
associated with the displacements decreased by a factor of more than 100.
Most of this decrease occurred during the first five iterations, after which
improvement was insignificant. The final value of $\chi^2$ is lower for this
optimization than in the verifications (Fig.~\ref{fig:verification}c) by about 
an order of magnitude; 
however, this is primarily due to the fact that the target parameters 
$\mathbf{X}^*$ determined here from the experimental data have greater
uncertainty than those of the manufactured solutions.

\begin{figure*}
    \begin{center}
    \includegraphics[width=0.9\textwidth]{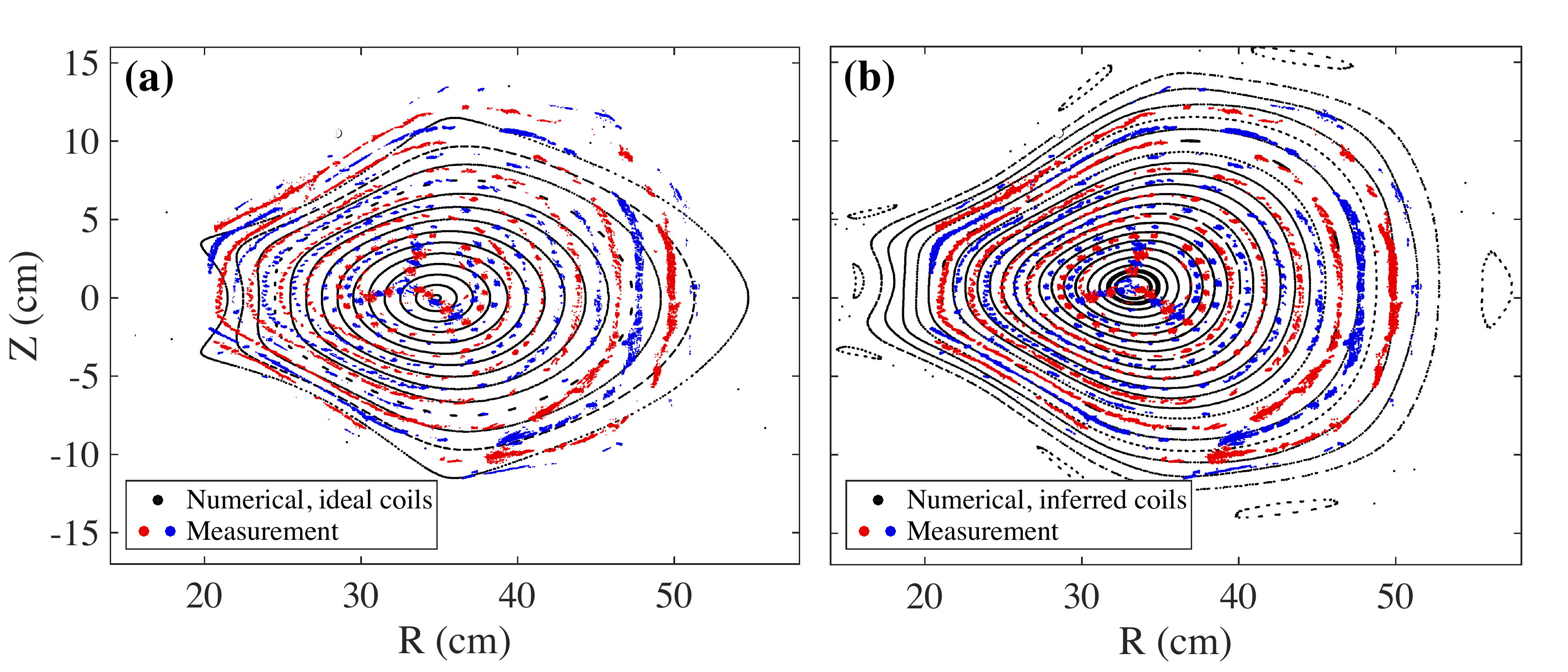}
    \caption{(a) Experimental data (red and blue dots) plotted alongside
             numerical data (black) for a field line trace that assumed no coil 
             displacements, both at $I_{IL}/I_{PF}=3.68$, identical to what 
             is plotted in Fig.~\ref{fig:ideal_poinc_comp}a. (b) The same 
             experimental data as in (a) plotted against numerical data for the
             coil displacements inferred from the optimization 
             ($\mathbf{p}^*_{3.68}$), again at $I_{IL}/I_{PF}=3.68$.}
    \label{fig:opt_poinc_comp}
    \end{center}
\end{figure*}

The qualitative improvement in the numerically predicted Poincar\'{e} plots 
generated using $\mathbf{p}^*_{3.68}$ versus $\mathbf{p}_0$ is shown in
Fig.~\ref{fig:opt_poinc_comp}.
Fig.~\ref{fig:opt_poinc_comp}a, identical to Fig.~\ref{fig:ideal_poinc_comp}a,
refers to $I_{IL}/I_{PF} = 3.68$ and compares the experimental data to a 
numerical prediction generated from $\mathbf{p}_0$ (\textit{i.e.}, coil 
displacements were neglected).
Note, again, the disagreements in the vertical elongation on the 
outboard side, the concavity on the inboard side, and in the position of the 
magnetic axis. Most of the numerical cross-sections intersect multiple 
experimental ones. 

Fig.~\ref{fig:opt_poinc_comp}b  shows the same experimental data overlaid on 
numerical data computed from $\mathbf{p}^*_{3.68}$.
The incorporation of the displacements 
in the comparison has dramatically improved the qualitative agreement. 
The discrepancies in magnetic axis, as well as the shaping of the inboard and 
outboard sides, has essentially vanished, and there far fewer cases of 
numerical cross-sections intersecting multiple experimental ones.

\begin{figure}
    \includegraphics[width=0.5\textwidth]{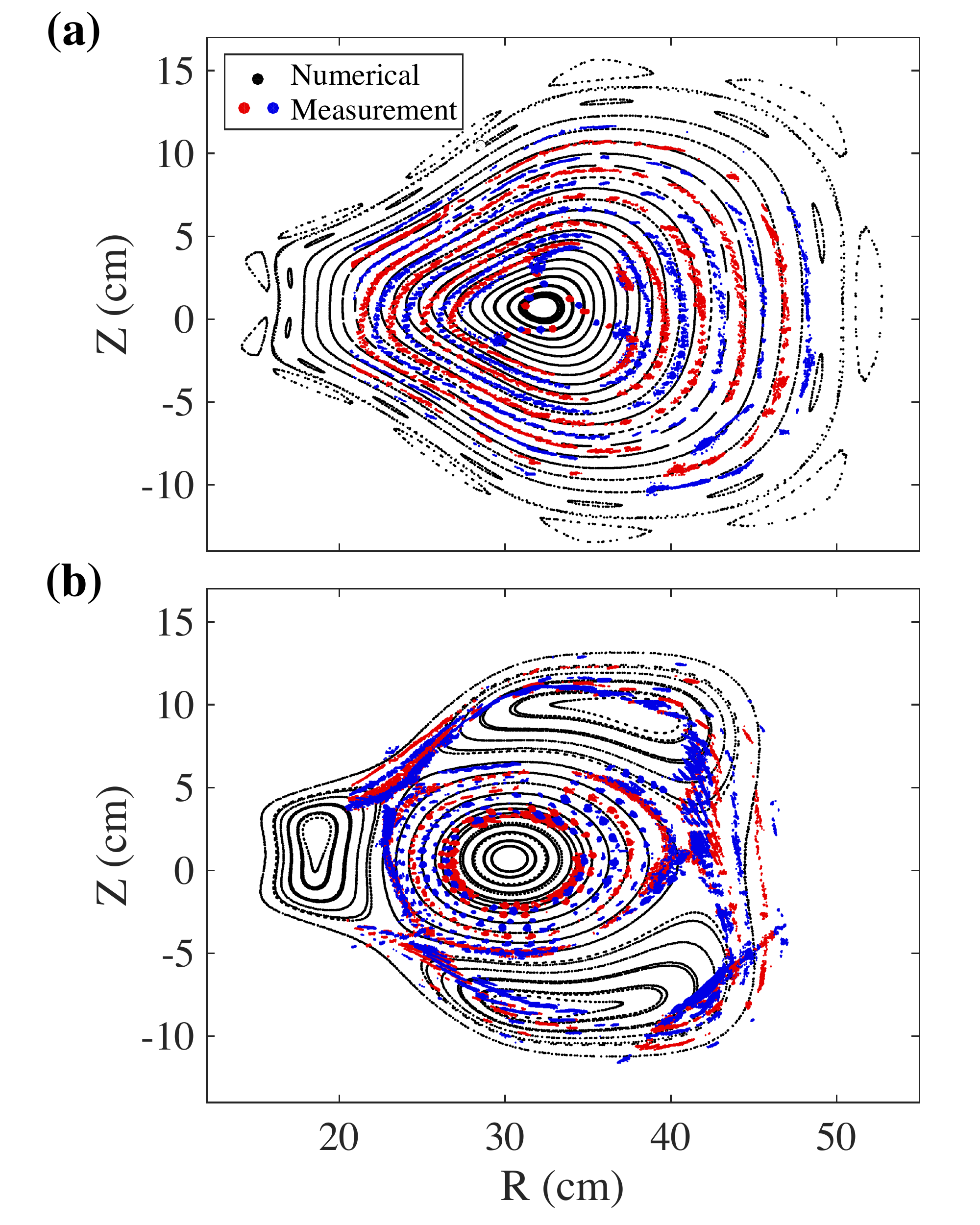}
    \caption{Effect of using, in calculations for (a) $I_{IL}/I_{PF}=3.50$
             and (b) $I_{IL}/I_{PF}=3.18$, the coil displacements inferred
             for $I_{IL}/I_{PF}=3.68$ (\textit{i.e.}, $\mathbf{p}^*_{3.68}$).
             Measurements for the respective current ratios
             are shown for comparison.}
    \label{fig:opt_poinc_otherRatios}
\end{figure}

Figs.~\ref{fig:opt_poinc_otherRatios}a-b show comparisons at other 
current-ratios ($I_{IL}/I_{PF} = 3.50$ for \ref{fig:opt_poinc_otherRatios}a and
$I_{IL}/I_{PF} = 3.18$ for \ref{fig:opt_poinc_otherRatios}b)
in which the numerical Poincar\'{e} data were generated from 
$\mathbf{p}^*_{3.68}$. The agreement is not as good as in 
Fig.~\ref{fig:opt_poinc_comp}b, which is to be expected because the error 
vector $\mathbf{F}$ used in the optimization was based exclusively on the 
geometry of the $I_{IL}/I_{PF}=3.68$ cross-section. It is noteworthy, however, 
that for $I_{IL}/I_{PF}=3.18$ the optimization nonetheless predicts three 
islands near the edge of size and position comparable to what is observed.
Note also the improvement in agreement for $I_{IL}/I_{PF}=3.18$ using 
$\mathbf{p}^*_{3.68}$ in Fig.~\ref{fig:opt_poinc_otherRatios}b over 
Fig.~\ref{fig:ideal_poinc_comp}b in which the numerical data were calculated 
from the nominal coil positions.

\begin{figure*}
    \begin{center}
    \includegraphics[width=0.9\textwidth]{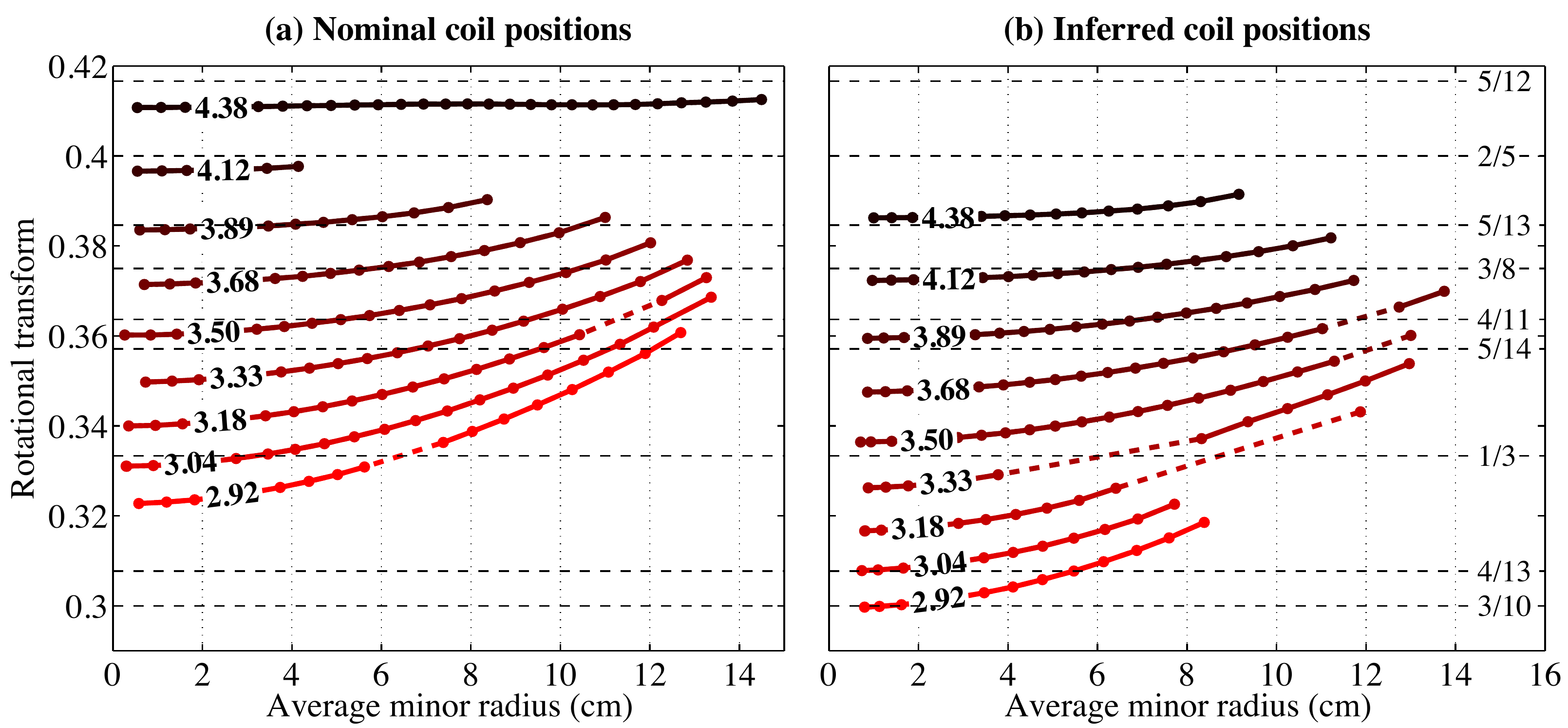}
    \caption{Comparison of \sout{$\iota$} profiles calculated for selected 
             current-ratios in the nominal 
             and optimized coil configurations. 
             (a) Profiles for the nominal IL coil positions ($\mathbf{p}_0$) 
             and assuming no PF coil displacements, identical to what is 
             plotted in Fig.~\ref{fig:iota_unp}.
             (b) Profiles computed using the optimized IL coil configuration
             ($\mathbf{p}^*_{3.68}$) and the measured PF coil displacements.
             Each dot represents a closed flux surface, so while the curves for 
             $I_{IL}/I_{PF}$ = 3.04 and 2.92 in plot (b) do not pass 
             through \sout{$\iota$} = 1/3, they contain large three-island 
             chains on their edges. Note that, in the optimized configuration 
             (b), current-ratios $I_{IL}/I_{PF}$ less than 3.5 contain
             \sout{$\iota$} = 1/3, consistent with observations.}
    \label{fig:iota_opt}
    \end{center}
\end{figure*}

Finally, calculated values of \sout{$\iota$} at various current-ratios 
$I_{IL}/I_{PF}$ using $\mathbf{p}^*_{3.68}$ are shown in 
Fig.~\ref{fig:iota_opt}b. Note that, for $I_{IL}/I_{PF} < 3.5$, the profiles of
\sout{$\iota$} contain the value 1/3, which is consistent with the observations
described in Sec.~\ref{subsect:exp_results}. Recall that the \sout{$\iota$} 
profiles that had been predicted for the nominal coil positions 
(Fig.~\ref{fig:iota_unp}) failed to account for the presence of 
\sout{$\iota$} = 1/3 for $I_{IL}/I_{PF} \geq 3.18$ and that, furthermore, the PF
coil displacements alone were insufficient to explain this disagreement 
(Fig.~\ref{fig:pfc_disp_compare}). Hence, the optimization of the IL coil 
displacements has resolved the discrepancy between the observations and 
predictions.

\section{Discussion}
\label{sect:discussion}

\subsection{Optimizing for cross-sections with islands}
\label{subsect:islands}

For the optimization conducted for the CNT IL coils in this paper, the target 
parameters in $\mathbf{X^*}$ came exclusively from a single current-ratio
topology ($I_{IL}/I_{PF}$ = 3.68) that had no significant magnetic islands. In 
principle, a cross-section containing magnetic islands should also be 
useable for an optimization if adequate experimental surface data are 
available, in the sense that many puncture points are
distributed evenly around the cross-section of the surface. Such data were not 
available from CNT cross-sections featuring prominent islands, largely as a 
result of two factors. 

The first was that cross-sections 
containing the large \sout{$\iota$} = 1/3 island chain were cut off at the 
inboard side due to the finite extent of the fluorescent rod. On the other
hand, had the rod extended further inboard, it would have collided with the IL 
coils during rotation. Thus, as seen in Fig.~\ref{fig:opt_poinc_otherRatios}b, 
there are no complete surfaces recorded outside the island chain. This factor 
may be unique to CNT as it relates to the particular coil configuration. 

The second factor, more likely to arise in other stellarators,
is the shadowing of near-rational surfaces in the vicinity of island chains. 
This is a result of the electron beam striking the back side of the 
gun before reaching the fluorescent rod. This effect is visible in surfaces 
near the magnetic axis in Fig.~\ref{fig:opt_poinc_comp}a-b, in which only three 
dots appear, corresponding to the first three toroidal transits 
of the electron beam. In this case,
the shadowing was not a major concern because only a small portion of the 
cross-section was affected. But in low-shear profiles with one 
or more rational surfaces at intermediate minor radii, large portions of the 
cross-section may be shadowed, thereby obscuring a large amount of geometric
information and preventing the determination of accurate 
geometric parameters.

However, if the above factors are absent or limited in extent, we 
expect that cross-sections with island chains should be useable for 
optimization. 

Poincar\'{e} data from field lines within a chain of large 
islands, identifiable by large empty intervals in the poloidal angle $\theta$, 
are ignored in the parametrization procedure (\ref{subsect:island-param}). 
Thus, only the closed flux surfaces on either side of the island chain will
determine $\mathbf{X}$, and, in turn, the Netwon steps $\delta\mathbf{p}$ in 
an optimization. Although the information
about the magnetic geometry within the islands themselves is lost, 
the kinking of nearby closed flux surfaces adjacent to the island chain is 
retained. It is hypothesized that the closed flux surfaces with 
island-induced kinking should be sufficient for the optimization.

\subsection{Extension to more complex devices}

Although the vector $\mathbf{p}$ of coil parameters used for the optimization 
in this paper had only ten components, this vector can in principle be expanded
to contain arbitrarily many parameters characterizing non-rigid displacements 
for multiple coils. One logical extension for CNT would be to allow for 
elliptical compression of the IL coils, which would add two more parameters for each coil (one for ellipticity and one for phase). More general deformations 
may be treated, for example, as Fourier series in which each additional $m$ 
number would require the addition of two parameters (coefficients)
if the coil is assumed to stay planar. With the addition of even more 
parameters, the coil could be allowed to deform off of its nominal plane.

It follows that, for an arbitrary stellarator, the size of $\mathbf{p}$ would 
scale as the number of coils to be optimized times the number of degrees of 
freedom for rotation, translation and deformation afforded to each coil. Of 
course, as seen in the foregoing study, each coil need not be given the same 
number of degrees of freedom.

Assuming $\mathbf{X}$ consists of geometric data from only one current-ratio
(which need not be the case), each column of the Jacobian can be computed with 
a single call to the field line tracer. Different columns, however,
corresponding to perturbations to different components of $\mathbf{p}$, 
require separate field line traces. Since the field line traces are the most
demanding part of the optimization procedure, the computation will 
scale linearly with the size of $\mathbf{p}$. It should be noted, however, that 
since the columns of the Jacobian may be determined in parallel with one
another, the time need not scale linearly if multiple processors are 
available.

\section{Summary and future work}
\label{sect:summary}

In summary, significant vacuum field errors have been observed in 
the CNT stellarator with the coils configured to have 
$\theta_{tilt}=78^\circ$. Photogrammetric measurements of the PF coils 
showed misalignments, although those misalignments were not sufficient to fully
explain the disagreements, in particular the observed offset in the 
\sout{$\iota$} profile. A numerical analysis of the influences of coil 
displacements on \sout{$\iota$} indicate that \sout{$\iota$} is most sensitive
to displacements of the IL coils. This motivated the development of a numerical
optimization method to calculate displacements for the IL coils that, in 
combination with the measured PF coil displacements, fit to the 
observed Poincar\'{e} cross-sections. The application of this algorithm 
to CNT at $I_{IL}/I_{PF}=3.68$ led to a set of IL coil displacements that 
exhibited significantly improved quantitative and qualitative agreement with 
observations.

Future work will focus on improving the optimization algorithm to achieve even
better fits to the observed data. These improvements will include (1) 
simultaneously optimizing to data from multiple current-ratios using 
$\mathbf{F}$ vectors that include geometric parameters for multiple Poincar\'{e}
cross-sections, (2) operating the field line tracer at greater numerical 
precision to reduce the uncertainty in numerically generated Fourier 
coefficients and allow for smaller finite difference intervals for the 
computation of the Jacobian, (3) expanding the $\mathbf{p}$ vector to 
include other sources of error including coil deformations, uncompensated coil 
leads, and displacements of the PF coils, and (4) investigating ways 
of generalizing the $\mathbf{X}$ vector to include information about 
Poincar\'{e} data inside islands, possibly analogous to generalizations of 3D 
toroidal equilibria found in codes like PIES \cite{rieman1986}, 
SIESTA \cite{hirshman2011}, and SPEC \cite{hudson2012}. 
Many of these improvements will
be more demanding computationally, although much of the algorithm can be
parallelized (in particular, the calculation of the covariance and Jacobian 
matrices).

While the misalignments in the CNT coils could in principle be corrected by,
for example, repairing or replacing the support structures, this is not a high
priority for the CNT program. One reason for this is that, in spite of the 
field errors, we still have access to configurations with good sets of nested, 
closed flux surfaces such as the one attained for $I_{IL}/I_{PF}$ = 3.68. In 
addition, the large island chains that the field errors have given rise to at 
lower current-ratios present an opportunity for research in island divertor 
physics.

\section{Acknowledgments}

The authors would like to thank S.~Lazerson for assistance with the 
field line tracing code employed in this study, as well as R.~Diaz-Pacheco and 
Y.~Wei for their assistance with data collection. The authors would also like to 
acknowledge the financial support of the Department of Energy and the 
National Science Foundation of the United States, Grant 
No.~NSF-PHY-04-49813.

\section*{References}
%\bibliography{bfield_2016.bib}

\begin{thebibliography}{29}

%\begin{thebibliography}{10}
\expandafter\ifx\csname url\endcsname\relax
  \def\url#1{{\tt #1}}\fi
\expandafter\ifx\csname urlprefix\endcsname\relax\def\urlprefix{URL }\fi
\providecommand{\eprint}[2][]{\url{#2}}

\bibitem{reimerdes2011}
Reimerdes H, Buttery R~J, Garofalo A~M, In Y, Haye R~J~L, Lanctot M~J,
  Okabayashi M, Park J~K, Schaffer M~J, Strait E~J and Volpe F~A 2011 {\em
  Fusion Science and Technology\/} {\bf 59} 572--585

\bibitem{hender2007}
Hender T~C, Wesley J~C, Bialek J, Bondeson A, Boozer A~H, Buttery R~J, Garofalo
  A, Goodman T~P, Granetz R~S, Gribov Y, Gruber O, Gryaznevich M, Giruzzi G,
  G\"{u}nther S, Hayashi N, Helander P, Hegna C~C, Howell D~F, Humphreys D~A,
  Huysmans G~T~A, Hyatt A~W, Isayama A, Jardin S~C, Kawano Y, Kellman A, Kessel
  C, Koslowski H~R, Haye R~J~L, Lazzaro E, Liu Y~Q, Lukash V, Manickam J,
  Medvedev S, Mertens V, Mirnov S~V, Nakamura Y, Navratil G, Okabayashi M,
  Ozeki T, Paccagnella R, Pautasso G, Porcelli F, Pustovitov V~D, Riccardo V,
  Sato M, Sauter O, Schaffer M~J, Shimada M, Sonato P, Strait E~J, Sugihara M,
  Takechi M, Turnbull A~D, Westerhof E, Whyte D~G, Yoshino R, Zohm H, the ITPA
  MHD~Disruption and Group M~C~T 2007 {\em Nuclear Fusion\/} {\bf 47}
  S128--S202

\bibitem{mynick2006}
Mynick H~E 2006 {\em Physics of Plasmas\/} {\bf 13} 058102

\bibitem{xanthopoulos2014}
Xanthopoulos P, Mynick H~E, Helander P, Turkin Y, Plunk G~G, Jenko F,
  G\"{o}rler T, Told D, Bird T and Proll J~H~E 2014 {\em Physical Review
  Letters\/} {\bf 113} 155001

\bibitem{sinclair1970}
Sinclair R~M, Hosea J~C and Sheffield G~V 1970 {\em Applied Physics Letters\/}
  {\bf 17} 92

\bibitem{colchin1989}
Colchin R~J, Anderson F~S~B, England A~C, Gandy R~F, Harris H~H, Henderson M~A,
  Hillis D~L, Kindsfather R~R, Lee D~K, Million D~L, Murakami M, Neilson G~H,
  Saltmarsh M~J and Simpson C~M 1989 {\em Review of Scientific Instruments\/}
  {\bf 60} 2680--2689

\bibitem{jaenicke1993}
Jaenicke R, Ascasibar E, Grigull P, Lakicevic I, Weller A, Zippe M, Hailer H
  and Schw{\"o}rer K 1993 {\em Nucl. Fusion\/} {\bf 33} 687--704

\bibitem{pedersen_pop2006}
Pedersen T~S, Kremer J~P, Lefrancois R~G, Marksteiner Q, Sarasola X and Ahmad N
  2006 {\em Phys. Plasmas\/} {\bf 13} 012502

\bibitem{colchin_iaea1989}
Colchin R~J, Harris J~H, Anderson F~S~B, England A~C, Gandy R~F, Hanson J~D,
  Henderson M~A, Hillis D~L, Jernigan T~C, Lee D~K, Lynch V~E, Murakami M,
  Neilson G~H, Rome J~A, Saltmarsh M~J and Simpson C~M 1989 Correction of field
  errors in the atf torsatron {\em Proceedings of the 16th European Conference
  on Controlled Fusion and Plasma Physics, Venice\/} ({\em Europhysics
  Conference Abstracts\/} vol 13B) (IAEA) pp 615--618

\bibitem{otte2003}
Otte M, Lingertat J and Wagner F 2003 Magnetic flux surface measurements with
  vertical field and compensation coils at the wega stellarator {\em
  Proceedings of the 14th International Stellarator Workshop, Greifswald\/}

\bibitem{sakakibara2013}
Sakakibara S, Narushima Y, Okamoto M, Watanabe K~Y, Suzuki Y, Ohdachi S, Ida K,
  Yoshinuma M, Tanaka K, Tokuzawa T, Narihara K, Yamada I, Yamada H and the LHD
  Experiment~Group 2013 {\em Nuclear Fusion\/} {\bf 53} 043010

\bibitem{hirsch2008}
Hirsch M, Baldzuhn J, Beidler C, Brakel R, Burhenn R, Dinklage A, Ehmler H,
  Endler M, Erckmann V, Feng Y, Geiger J, Giannone L, Geiger G, Grigull P,
  Hartfuss H~J, Hartmann D, Jaenicke R, K\"{o}nig R, Laqua H~P, Maassberg H,
  McCormick K, Sardei F, Speth E, Stroth U, Wagner F, Weller A, Werner A, Wobig
  H, Zoletnik S and the W7-AS~Team 2008 {\em Plasma Physics and Controlled
  Fusion\/} {\bf 50} 053001

\bibitem{rummel2012}
Rummel T, Risse K, Kisslinger J, K\"{o}ppen M, F\"{u}llenback F, Neilson H,
  Brown T and Ramakrishnan S 2012 {\em IEEE Transactions on Applied
  Superconductivity\/} {\bf 22} 4201704

\bibitem{kisslinger2005}
Kisslinger J and Andreeva T 2005 {\em Fusion Engineering and Design\/} {\bf 74}
  623--626

\bibitem{landreman2016}
Landreman M and Boozer A~H 2016 {\em Phys. Plasmas} {\bf 23} 032506

\bibitem{pedersen2004}
Pedersen T~S, Boozer A~H, Kremer J~P, Lefrancois R~G, Reiersen W~T, Dahlgren
  F~D and Pomphrey N 2004 {\em Fusion Sci. Technol.\/} {\bf 46} 200

\bibitem{imagawa1998}
Imagawa S, Masuzaki S, Yanagi N, Yamaguichi S, Satow T, Yamamoto J, Motojima O
  and the LHD~group 1998 {\em Fusion Engineering and Design\/} {\bf 41} 253

\bibitem{wanner2000}
Wanner M and the W7-X~team 2000 {\em Plasma Physics and Controlled Fusion\/}
  {\bf 42} 1179

\bibitem{lazerson2016}
Lazerson S, Otte M, Bozhnekov S, Biedermann C, Pedersen T~S and the W7-X~team
  2016 {\em Nuclear Fusion (submitted)\/}

\bibitem{brenner2008}
Brenner P~W, Pedersen T~S, Berkery J~W, Marksteiner Q~R and Hahn M~S 2008 {\em
  IEEE Transactions on Plasma Science\/} {\bf 36} 1108

\bibitem{kremer_thesis}
Kremer J~P 2006 {\em The creation and first studies of electron plasmas in the
  Columbia Non-neutral Torus\/} Ph.D. thesis Columbia University New York, NY
  10027

\bibitem{lao1985}
Lao L~L, John H~S, Stambaugh R~D, Kellman A~G and Pfeiffer W 1985 {\em Nuclear
  Fusion\/} {\bf 25} 1611

\bibitem{hanson2009}
Hanson J~D, Hirshman S~P, Knowlton S~F, Lao L~L, Lazarus E~A and Shields J~M
  2009 {\em Nuclear Fusion\/} {\bf 49} 075031

\bibitem{press1992}
Press W~H, Teukolsky S~A, Vetterling W~T and Flannery B~P 1992 {\em Numerical
  Recipes in C\/} 2nd ed (Cambridge University Press)

\bibitem{jones2006}
Jones C~S and Finn J~M 2006 {\em Nuclear Fusion\/} {\bf 46} 335

\bibitem{islr2013}
James G, Witten D, Hastie T and Tibshirani R 2013 {\em An Introduction to
  Statistical Learning with Applications in R\/} (Springer)

\bibitem{rieman1986}
Rieman A and Greenside H 1986 {\em Computer Physics Communications\/} {\bf 43}
  157

\bibitem{hirshman2011}
Hirshman S~P, Sanchez R and Cook C~R 2011 {\em Physics of Plasmas\/} {\bf 18}
  062504

\bibitem{hudson2012}
Hudson S~R, Dewar R~L, Dennis G, Hole M~J, McGann M, von Nessi G and Lazerson S
  2012 {\em Physics of Plasmas\/} {\bf 19} 112502

\end{thebibliography}

\appendix

\section{Parametrization of Poincar\'{e} cross-sections}
\label{sect:parametrization}

The level of agreement between two sets of Poincar\'{e} data is determined by 
fitting the nested flux surface cross-sections to a discrete set of geometric
parameters $\mathbf{X}$. The components of $\mathbf{X}$ are the coefficients 
of a linear combination of orthogonal functions consisting of a Fourier 
series in the poloidal angle $\theta$ and a polynomial series in the normalized
minor radius $\rho$. 

For a flux surface 
characterized by some particular $\rho$, the $R$ and $Z$ coordinates 
of the point with poloidal angle $\theta$ are expressed as 

\begin{equation*}
    R(\rho, \theta) = R_0(\rho) + \sum_{m=1}^{M}R_{cm}(\rho)\cos(m\theta) 
\end{equation*}
\begin{equation}
    \phantom{R(\rho, \theta) =} + \sum_{m=1}^{M}R_{sm}(\rho)\sin(m\theta) 
\end{equation}
\begin{equation*}
    Z(\rho, \theta) = Z_0(\rho) + \sum_{m=1}^{M}Z_{cm}(\rho)\cos(m\theta)
\end{equation*}
\begin{equation}
    \phantom{R(\rho, \theta) =} + \sum_{m=1}^{M}Z_{sm}(\rho)\sin(m\theta) 
\label{eqn:fourier}
\end{equation}

As indicated in the above representation, the coefficients of each Fourier 
mode are themselves functions of $\rho$. This dependence is represented as a
linear combination of polynomials $P_s(\rho)$; for example:

\begin{equation}
    R_{c1}(\rho) = R_{c10}P_0(\rho) + R_{c11}P_1(\rho) + ... + 
                   R_{c1S}P_S(\rho)
\label{eqn:polyCoeffs}
\end{equation}

\noindent The polynomials $P_s(\rho)$ are chosen to be orthonormal in the inner 
product defined by

\begin{equation}
\langle P_i, P_j \rangle = \int_0^1 P_i(\rho)P_j(\rho)d\rho = \delta_{ij}, 
\end{equation}

\noindent and the first few polynomials in this set are listed in Table 
\ref{tab:polynomials}.

\begin{table}
\begin{tabular}{|l|l|}
    \hline
    $s$ & $P_s(\rho)$ \\
    \hline
    $0$ & $1$ \\
    $1$ & $\sqrt{3}\left(2\rho - 1\right)$ \\
    $2$ & $\sqrt{5}\left(6\rho^2 - 6\rho + 1\right)$ \\
    $3$ & $\sqrt{7}\left(20\rho^3 - 30\rho^2 + 12\rho - 1\right)$ \\
    $4$ & $      3 \left(70\rho^4 - 140\rho^3 + 90\rho^2 - 20\rho + 1\right)$ \\
    \hline
\end{tabular}
\caption{The first few polynomials $P_s(\rho)$ as discussed in the text.}
\label{tab:polynomials}
\end{table}

The vector $\mathbf{X}$ of geometric parameters, then, is just a list of these
coefficients: 

\begin{equation*}
    \mathbf{X} = 
        \{ 
            R_{00},~...,~R_{0S},~R_{c10},~...,~R_{c1S},~...,~R_{cMS},~ 
        \phantom{\}}
\end{equation*}
\begin{equation}
    \phantom{\mathbf{X} = \{}
            R_{s01},~R_{sMS},~Z_{00},~...,~Z_{sMS}
        \}
\end{equation}

To ensure that the set of coefficients is unique for a particular set of 
Poincar\'{e} data, it is necessary to use a precise definition for $\rho$ and 
$\theta$. In this work, $\rho$ for a particular cross-section is defined as 
$\sqrt{A/A_{max}}$, where $A$ is the enclosed area and $A_{max}$ is 
a reference area not to exceed the area enclosed by the last closed flux 
surface. In this way, $\rho=0$ at the magnetic axis and $\rho=1$ on the edge of 
the region of the Poincar\'{e} cross-section to be parametrized. The poloidal 
angle $\theta$ is defined to advance in direct proportion to the arc length 
along a flux surface cross-section and is set to zero on the outboard side 
where $Z$ is equal to $Z(\rho=0)$. 

\section{Determination of the $\mathbf{X}$ vector for Poincar\'{e} data}
\label{sect:coeff_determination}

\begin{figure}
    \includegraphics[width=0.5\textwidth]{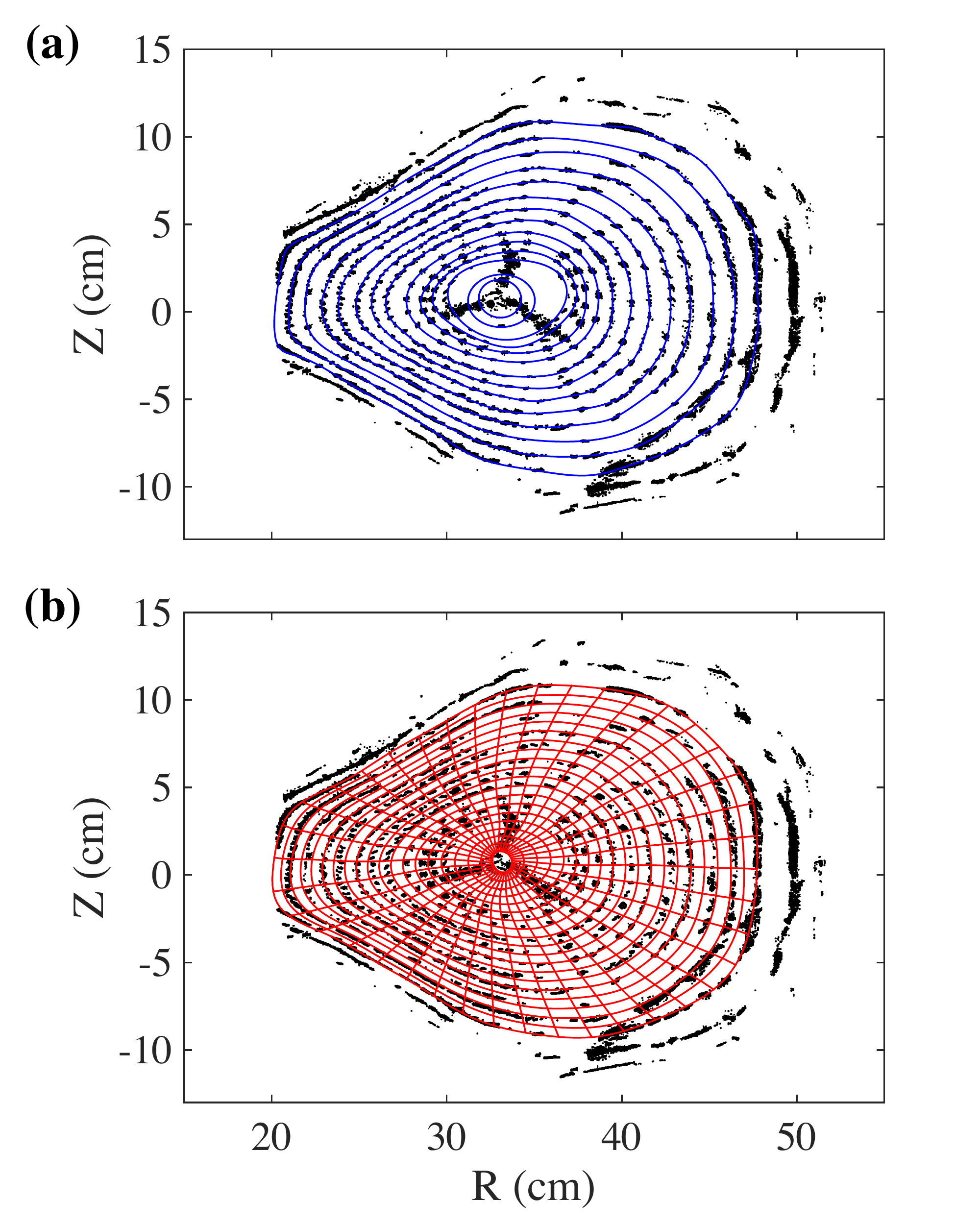}
    \caption{(a) Experimental Poincar\'{e} data obtained for the current-ratio 
             3.68 (black dots) and fit curves to selected flux surfaces 
             (blue lines). (b) The same experimental data overlaid with 
             level curves from the fitted geometric parameters in red. 
             Radially extending red lines are 
             curves of constant $\theta$; closed red loops are curves of 
             constant $\rho$.}
    \label{fig:data2params}
\end{figure}

An example of a fit of coefficients to experimental data is shown in 
Fig.~\ref{fig:data2params}. As discussed in Section \ref{sect:flux_surf_meas}, 
the experimental data were separated according to flux surface. For each 
flux surface, a curve was fit iteratively to the Poincar\'{e} points in such 
a way as to minimize the disagreement with the data points. An example of a set 
of fit curves is shown in Fig.~\ref{fig:data2params}a. When the fitting 
iterations are complete, the curve was parametrized in $\theta$ as described 
in \ref{sect:parametrization} and the $i^{th}$ measured surface was assigned a 
$\rho_i$ value based on 
the enclosed area. Each fit curve 
$\left(R\left(\rho_i,\theta\right), \: Z\left(\rho_i,\theta\right)\right)$
was then projected onto a Fourier series, yielding one set of coefficients 
$\{R_{cm}\left(\rho_i\right), \: R_{sm}\left(\rho_i\right), \:
   Z_{cm}\left(\rho_i\right), \: Z_{cm}\left(\rho_i\right)\}_{m=0}^M$
for each value $\rho_i$.
Sets of corresponding Fourier coefficients (e.g.,
$\{R_{c1}\left(\rho_i\right)\}_i$) were then fit by linear least-squares onto 
polynomials in $\rho$ of degree $S$, from which the polynomial coefficients as 
shown in Eq.~\ref{eqn:polyCoeffs} were obtained. The calculations described 
in this paper used $M = 14$ and $S = 7$.

Because experimental data points are often limited near the magnetic axis, 
certain constraints are enforced in the least-squares fitting in order to 
avoid spurious oscillations in the $\rho$ polynomials. In particular, all 
polynomials for Fourier coefficients of $m \ge 1$ are constrained to be zero 
at the axis. In addition, all coefficients of $m \ge 2$ are constrained to have a first derivative of zero at the axis. Furthermore, coefficients with $m = 0$ 
are only expanded to second order in $\rho$ (\textit{i.e.}, requiring $s\leq2$).

Once coefficients are calculated for a set of Poincar\'{e} data, they are 
arranged into a vector $\mathbf{X}$ containing $(S+1)(4M+2)$ elements. 
Because of the constraints described in the above paragraph, $2(S-2) + 
8M - 4$ of those components are redundant, and are therefore removed 
for subsequent analysis.

Ideally, the coefficients $\mathbf{X}$ should be unique for a given set of 
Poincar\'{e} data. However, in practice the coefficients will vary slightly 
depending on which Poincar\'{e} points are used for this fit. Specifically, 
for experimental data, this depends on which surfaces are measured and which 
of the subsequent pixels are used for the analysis. For numerical data, the 
dependence is on the locations where the field line traces are initialized.
Hence, each component of $\mathbf{X}$ will have an associated uncertainty that 
may be correlated with that of other components. 
Sec.~\ref{subsect:opt_proc} and \ref{subsect:numerical} describe how 
these correlations are accounted for in the optimization.

\subsection{Treatment of magnetic islands}
\label{subsect:island-param}

If a chain of large islands exists within a cross-section, it will be identified
during the curve-fitting process by poloidal gaps in the Poincar\'{e} data.
Since the parametrization described in \ref{sect:parametrization} only permits
closed flux surfaces, the islands are simply ignored during in the fits of 
Fourier coefficients to the polynomials $P_s(\rho)$ decribed above. 
Since some Fourier coeffients may change abruptly from the axis-facing side
of the island to the edge-facing side, it may be
necessary to incorporate higher-order polynomials $P_s(\rho)$ in the fits, 
leading to more components in the $\mathbf{X}$ vector.

If one were to plot the curves of constant $\rho$
specified from the geometric parameters $\mathbf{X}$ (as in 
Fig.~\ref{fig:data2params}b), one would observe a continuous deformation
from the shape of the last closed flux surface on the 
axis-facing side of the island chain to the first closed surface on the 
edge-facing side. These deformed curves in the island region clearly do not 
reflect the actual 
island geometry and are simply a result of interpolation between the data for
the core-facing side and the edge-facing side. In other words, the laminar
flux surfaces occurring in place of the islands reflect the fact that the 
$\mathbf{X}$ vector contains information only about the flux surfaces on 
either side of the island chain.

\end{document}